%% file: attenuation_letter.tex
\documentclass{emulateapj}

\usepackage{ifthen} \usepackage{comment} 

\newboolean{includeversions} \setboolean{includeversions}{false}

\newboolean{blackandwhite} 

\newboolean{usepdf} \newif\ifpdf \ifx\pdfoutput\undefined \pdffalse \else \pdfoutput=1 \pdftrue \fi

\ifpdf \setboolean{usepdf}{true} \else \setboolean{usepdf}{false} \fi

\ifthenelse {\boolean{usepdf}}{ 
\ifthenelse{\boolean{blackandwhite}}{ \usepackage[pdftex,final, pdftitle={Simulations of Dust in Interacting Galaxies}, pdfauthor={Patrik Jonsson}, pdfsubject={Ph.D. thesis}, pdfpagemode={UseOutlines}]{hyperref} }{ \usepackage[pdftex,final, pdftitle={Simulations of Dust in Interacting Galaxies}, pdfauthor={Patrik Jonsson}, pdfsubject={Ph.D. thesis}, pdfpagemode={UseOutlines}, colorlinks,linkcolor={blue},citecolor={blue},urlcolor={red}]{hyperref} } }{ 
\usepackage[dvips]{hyperref} } \graphicspath {{./}}

\include {natlatex_macros}

\slugcomment{Submitted to \apj.}

\shorttitle{Dust Attenuation in Simulations of Interacting Galaxies} \shortauthors{Jonsson et al.}

\begin{document}


\title{Simulations of Dust in Interacting Galaxies I: Dust Attenuation}


\author{Patrik Jonsson}  \affil{Department of Astronomy and Astrophysics, University of   California, Santa Cruz, CA 95064} \email{patrik@ucolick.org}

\author{T. J. Cox\altaffilmark{1} and Joel R. Primack} \affil{Department of Physics, University of California,   Santa Cruz, CA 95064} \email{tcox@cfa.harvard.edu} \email{joel@scipp.ucsc.edu}

\and

\author{Rachel S. Somerville} \affil{Space Telescope Science Institute, 3700 San Martin Drive, Baltimore MD 21218} \email{somerville@stsci.edu}


\altaffiltext{1}{Present address: Harvard-Smithsonian Center for Astrophysics, 60 Garden Street,   Cambridge, MA 02138}



\begin{abstract}
  A new Monte-Carlo radiative-transfer code, \mcrx , is used in  
conjunction with hydrodynamic simulations of major galaxy mergers to  
calculate the effects of dust in such systems. The simulations are   in
good agreement with observations of dust absorption in starburst  
galaxies, and the dust has a profound effect on their appearance.   The
dust attenuation increases with luminosity such that at peak  
luminosities $\sim 90 \%$ of the bolometric luminosity is   absorbed by
dust.  In general, the detailed appearance of the   merging event
depends on the stage of the merger and the geometry of   the encounter.
The fraction of bolometric energy absorbed by the   dust, however, is a
robust quantity that can be predicted from the   intrinsic properties
bolometric luminosity, baryonic mass,   star-formation rate, and
metallicity of the system.  This paper   presents fitting formulae,
valid over a wide range of masses and   metallicities, from which the
absorbed fraction of luminosity (and   consequently also the infrared
dust luminosity) can be predicted.   The attenuation of the luminosity
at specific wavelengths can also   be predicted, albeit with a larger
scatter due to the variation with   viewing angle.  These formulae for
dust attenuation appear to be   valid for both isolated and interacting
galaxies, are consistent   with earlier studies, and would be suitable
for inclusion in   theoretical models, e.g. semi-analytic models of
galaxy formation.
\end{abstract}




%

\section{Introduction}

Galaxy mergers produce some of the most spectacular events in the
universe.  Locally, they are responsible for the most luminous objects,
Ultraluminous Infrared Galaxies
\citep[ULIRGs,][]{sandersmirabel96}.
At higher redshifts they may be responsible for the sources seen in the
submillimeter \citep{smailetal97}, and they may even be a dominant mode
of star formation in the early universe
\citetext{\citealp{SPF,elbazcesarsky03}, but see \citealp{wolf04}}.
Because the most luminous objects are also generally the most
dust-obscured, it was not until the launch of IRAS that the existence
of ULIRGs was discovered.  What in the optical appeared to be fairly
unimpressive, albeit peculiar, galaxies turned out to be the brightest
infrared sources in the local universe.  It is now clear that
starbursts and dust generally go hand-in-hand.  The large amounts of
gas necessary to fuel a major starburst bring with them large column
densities of dust, obscuring the starburst and reradiating the energy
in the far infrared.  In addition, rapid star formation quickly
enriches the region with metals, increasing the amount of dust further.
Including the effects of dust is thus crucial when studying these
systems.

Theoretical studies of dust in galaxies have generally not leveraged
the power of hydrodynamic simulations but rather used fairly simple
geometries.  \citet{gcw97}, using a Monte-Carlo radiative-transfer
model, concluded that Small-Magellanic-Cloud-type (as opposed to
Milky-Way-type) dust was necessary to fit observations of starburst
galaxies, and that their spectral energy distribution (SED) was best
replicated by a model where a clumpy shell of dust surrounds the
starburst.  \citet{ferraraetal99} calculated the effect of dust in
various analytical disk plus bulge models, also using a Monte-Carlo
model.  \citet{silvaetal98} and \citet{charlotfall00} modeled dust
effects using a simple two-phase model, where young stars are located
in dense molecular clouds while older stars are extinguished only by
the diffuse ISM, and showed that their models could fit spectral energy
distributions of various observed galaxies.  However, all these models
had numerous free parameters and gave little guidance on how to choose
them, thus limiting their predictive capability.

Notable previous efforts at simulating merging galaxies using
hydrodynamic N-body codes with star formation and feedback are
\citet{mihoshernquist94ulirgs, mihoshernquist96} and
\citet{springel00}, but these efforts did not consider realistic
observations of their simulations including the effects of dust.
\citet{bekkishioya01,bekkishioya00a,bekkishioya00b} did use a simple
model for dust attenuation and reradiation along with N-body
simulations to investigate major mergers.  However, their modeling,
using the ``sticky particle'' method \citep{schwarz81} and with no
supernova feedback, did not allow them to capture essential features of
the hydrodynamics of merging galaxies.  This work builds on these
earlier works, using a full radiative-transfer model to study the
effects of dust in the most comprehensive and sophisticated suite of
major-merger simulations done so far.  The radiative-transfer model is
described in
\citet{pjthesis-nourl}\footnote{Available at \url{http://sunrise.familjenjonsson.org/thesis}.},
and details about the hydrodynamic simulations can be found in
\citet{tjthesis-nourl}\footnote{Available at \url{http://physics.ucsc.edu/~tj/work/thesis}.}
and \citet{coxetal05methods}.

In this paper, we emphasize one particular aspect of these simulations:
the presence of a tight correlation between the fraction of energy
absorbed and quantities such as the luminosity, metallicity, and mass
of the simulated systems. These relations should be useful for
theoretical modeling of dust absorption in galaxies, for example in
semi-analytic models of galaxy formation.  This paper is the first in a
series describing the effects of dust in the simulations, the
appearance of the simulations and comparisons to observed starburst
galaxies (Jonsson et al. and Lotz et al., in preparation) including
using new nonparametric measures of morphology \citep{lotzetal04gm20}. 
Other papers based on these simulations concern the creation of a halo
of hot, shock-heated gas around the merger remnant \citep{coxetal04},
the star-formation properties of minor as well as major merger
simulations (Cox et al., in preparation),

\section{Model Description}

The simulations, performed with the ``entropy-conserving'' version of
the GADGET SPH code \citep{springeletal01, springelhernquist02},
consist of encounters of identical copies of the galaxies listed in
Table~\ref{table_mass_metallicity}.  Because of the very large
parameter space, the number of simulations needed to exhaustively
sample all possible initial conditions would be prohibitively large (on
the order of thousands).  Instead, the strategy used was to define a
few reasonably realistic cases based on information from observations
and cosmological simulations.

The main differences from previous work are the inclusion of efficient
supernova feedback, a correct normalization of the star-formation law,
and an accurate treatment of hydrodynamics \citep{coxetal05methods}.
These served to make the merger-induced starbursts less intense.  Peak
star-formation rates for major mergers of typical late-type spiral
galaxies are generally in the range $30 - 50 \Msun \yr^{-1}$, except
for very short spikes.

\subsection{Model Galaxies}

Several different dynamically self-consistent disk-galaxy models,
listed in Table~\ref{table_mass_metallicity}, are used in this study.

The Sbc galaxy was modeled using median properties for Sbc galaxies
from \citet{robertshaynes94} and bulge information from
\citet{dejong96}.  Lower- and higher-mass variants of the Sbc galaxy,
called Sbc- and Sbc+, respectively, were modeled using the 25th and
75th percentiles of the distributions of properties.  Another set of
galaxies, the ``G-series'', was based on mean relations from the Sloan
Digital Sky Survey \citep{shenetal03}, gas properties from
\citet{belletal03},  and bulge information from \citet{dejong96}. 
These galaxies have significantly less gas than the Sbc models and
cover a much larger range in mass.  Detailed information about these
and other model galaxies can be found in \citet{tjthesis-nourl}. The
galaxies were assigned a metallicity consistent with
\citet{zaritskyetal94}, used for both the stellar SEDs and the initial
gas metallicity.  All in all, the seven galaxy models cover a range of
about 100 in baryonic mass, 3 in gas fraction, and 4 in metallicity.

\begin{deluxetable}{lccccc} \tablecaption{   The galaxy models used for the merger simulations.   \label{table_mass_metallicity}} \tablecolumns{6} \tablehead{   \colhead{Model} &    \colhead{$M_\mathrm{vir}$\tablenotemark{a}} &    \colhead{$M_\mathrm{b}$\tablenotemark{b}} &    \colhead{$f_\mathrm{g}$\tablenotemark{c}} &    \colhead{$V_{\mathrm{rot}}$\tablenotemark{d}} &    \colhead{Z\tablenotemark{e}} \\   &    \colhead{($\mathrm{M}_\odot$)} &    \colhead{($\mathrm{M}_\odot$)} &    &    \colhead{($\mathrm{km\,s}^{-1}$)} &    \colhead{(Z$_\odot$)} } \startdata Sbc+ & $9.28\cdot10^{11}$ & $1.56\cdot10^{11}$ & 0.58 & $210$ & $1.12$ \\  Sbc & $8.12\cdot10^{11}$ & $1.03\cdot10^{11}$ & 0.58 & $195$ & $1.00$ \\  G3  & $1.16\cdot10^{12}$ & $6.22\cdot10^{10}$ & 0.20 & $192$ & $1.00$ \\  Sbc- & $3.60\cdot10^{11}$ & $4.98\cdot10^{10}$ & 0.58 & $155$ & $0.70$ \\  G2   & $5.10\cdot10^{11}$ & $1.98\cdot10^{10}$ & 0.24 & $139$ & $0.56$ \\  G1  & $2.00\cdot10^{11}$ & $7.00\cdot10^{9\phn}$ & 0.29 & $103$ & $0.40$ \\  G0   & $5.10\cdot10^{10}$ & $1.60\cdot10^{9\phn}$ &  0.38& \phn$67$ & $0.28$ \\  \enddata \tablenotetext{1}{ Virial mass.} \tablenotetext{2}{ Baryonic mass.} \tablenotetext{3}{ Gas fraction (of baryonic mass).} \tablenotetext{4}{ Circular velocity.} \tablenotetext{5}{ Metallicity (gas and stellar).} \tablecomments{For the Sbc galaxies, 9 mergers with different orbital   geometries were used.  For the other galaxies, only a prograde-prograde   encounter was used.} \end{deluxetable} 

\subsection{Galaxy Merger Setup}

Identical pairs of each of the seven galaxies in
Table~\ref{table_mass_metallicity} were put on a parabolic orbit with
the disks prograde.  One of the disks was in the plane of the orbit,
while the other was tilted by $30 \degrees$.  Eight additional mergers
of the Sbc galaxies were also simulated, exploring variations in galaxy
orientation and encounter orbit.  Details about the merger setup can be
found in \citet{tjthesis-nourl}.

\subsection{Radiative-Transfer Model}

The Monte-Carlo radiative-transfer code \mcrx , similar to the DIRTY
code \citep{gordonetal01}, is described in detail in
\citet{pjthesis-nourl} and in a future paper. Briefly, the geometry of
gas and stars along with the detailed star-formation histories of the
galaxies at roughly 50 different points in time were taken from the
N-body simulations and used as inputs to the radiative-transfer
calculation.  Outputs are multi-wavelength images of the system from a
number of viewpoints, as well as luminosity absorbed by dust and
reradiated in the infrared. \mcrx \ is available to interested
prospective
users.\footnote{The Sunrise web site is   \url{http://sunrise.familjenjonsson.org}.}

The stellar SEDs were taken from the Starburst99
\citep[v4.0,][]{leithereretal99}
stellar population synthesis model.  The IMF used, similar to a Kroupa
IMF \citep{kroupa02}, has a high-mass slope of -2.35 from the upper
limit of $150 \Msun$ down to $1 \Msun$.  Below this mass it is flat
down to the lower limit of $0.1 \Msun$. The disk stars existing at the
start of the simulation were assumed to have been forming at a uniform
rate for the previous $8 \Gyr$.  The bulge stars were assumed to have
formed in an instantaneous burst $8 \Gyr$ earlier.

To be able to describe the simulation geometry with sufficient accuracy
while keeping computational requirements reasonable, \mcrx \ uses an
adaptive grid.  The grid refinement criteria are adjustable; the
parameters used are described in \citet{pjthesis-nourl}.  The number of
cells in the resulting grids ranged from around 70,000 at the start of
the simulation to almost 300,000 at the time of peak star-formation
rate and highest densities.

The dust grain model used was the ``$R = 3.1$'' Milky-Way model by
\citet{weingartnerdraine01} with a dust-to-metal ratio $m_d / m_m =
0.4$ \citep{dwek98}.  The radiative-transfer calculation was performed
for 22 different wavelengths between $21 \nm$ and $5 \um$, including
the $\halpha$ and $\hbeta$ nebular emission lines. One million
Monte-Carlo rays were traced for each wavelength.  Rather than
self-consistently computing the infrared dust-emission spectrum, the
templates of \citet{devriendtetal99}, interpolated to the correct
luminosity, were used.

\section{Results}

\begin{figure} \begin {center} \includegraphics[width=\columnwidth]{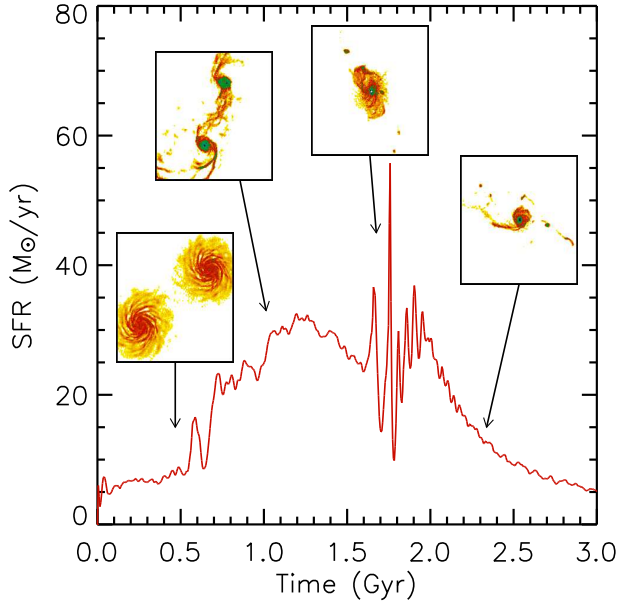} \end {center}  \caption{ \label{plot_fiducial_sfr} Overview of a major-merger event.  The graph shows star-formation rate as a function of time for the prograde-prograde merger between two Sbc galaxies.  The figures show the distribution of gas in the simulations at the indicated points in time from a viewpoint perpendicular to the orbital plane.  } \end{figure} \begin{figure} \begin {center} \includegraphics[width= 0.99\columnwidth]{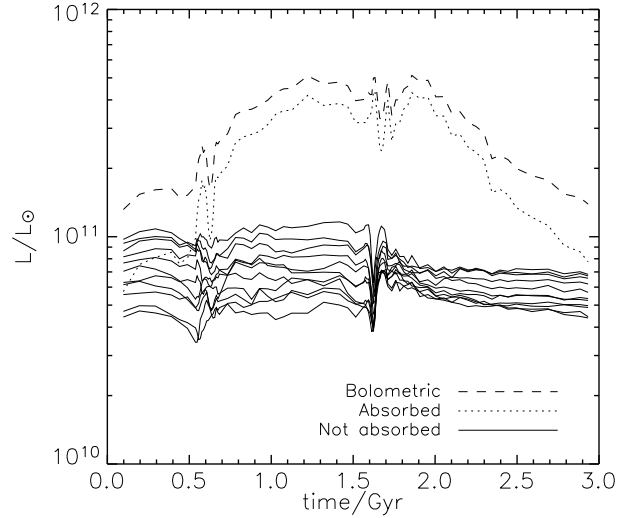} \end {center}  \caption{ \label{plot_fiducial_luminosity} Time evolution of bolometric luminosity and dust attenuation for the prograde-prograde Sbc merger simulation.  The bolometric luminosity of the system is shown with a dashed line, while the amount of luminosity absorbed by dust and reradiated in the infrared is shown with a dotted line.  The solid lines show the luminosity \emph{not} absorbed by dust, emerging in the UV/visual (more precisely the luminosity inferred by an observer under an assumption of isotropy), one line for each of the 11 viewing angles.  The luminosity emerging in the UV/visual is practically independent of the bolometric luminosity and shows significant variation with viewing angle.  } \end{figure}

First, some general results from the radiative-transfer simulations
will be presented.  Then we will continue with the specific results
concerning dust attenuation in the simulations. Throughout this paper,
``attenuation'' will be understood to refer to the fraction by which
luminosity is decreased due to dust, either averaged over all
directions or along a specific line of sight.  The term attenuation is
preferred over absorption to emphasize that both absorption and
scattering processes in a complex geometry contribute to a net decrease
(or, in rare cases, increase) in the emerging radiation.

\subsection{Overview of a Major Merger}

While the detailed behavior of a galaxy merger depends on the initial
setup of the two galaxies, the general sequence of events remains the
same.  Figure~\ref{plot_fiducial_sfr} shows the star-formation rate of
a merger between two Sbc galaxies, along with images of the system at
different points in time.  As the two galaxies initially approach each
other, they are largely unaffected by each other's presence.  Tidal
forces disrupt the galaxy disks at the time of the first close passage.
 The star-formation rate then increases due to gas being driven to the
centers as the galaxies separate.  The star-formation rate remains
elevated as the galaxies turn around and again approach.  There are
several short bursts of star formation as the two components merge. 
After coalescence, the merger remnant relaxes, and the star-formation
rate trails off over a timescale of about 1 Gyr.

The evolution of the bolometric luminosity of the same simulation is
shown in Figure~\ref{plot_fiducial_luminosity}.  The bolometric
luminosity of the system increases by almost an order of magnitude
during the course of the merger, closely following the star-formation
rate, as expected for a starburst system where young stars dominate the
energy output.  However, the fraction of luminosity which is absorbed
by dust scales with luminosity in such a way that the luminosity
\emph{not} absorbed by dust typically stays almost constant.  As the
star-formation rate and hence the luminosity increase, the column
densities of gas and dust also go up, and with them the attenuation;
the two factors appear to conspire to keep the UV/visual luminosity
roughly constant.  This effect can be seen in
Figure~\ref{plot_fiducial_luminosity}, where the solid lines, showing
the luminosity escaping in different directions, show little variation
with time.  This tendency was also observed by \citet{bekkishioya00a}
in their simulations, and in observational studies
\citep{adelbergersteidel00}.

While there is little temporal variation in the luminosity not absorbed
by dust, there is, in general, a substantial difference between the
flux escaping in different directions. Because the initial systems
consist of disk galaxies, radiation escapes preferentially out of the
plane of the galaxies.  This creates a difference of around a factor of
2 in the escaping flux, clearly visible in
Figure~\ref{plot_fiducial_luminosity}. The merger remnant is more
spheroidal and thus the difference with viewing angle is smaller at the
end stage of the mergers. As will be shown below, this variation in
viewing angle means that quantities not dependent on direction, such as
the fraction of bolometric luminosity which is absorbed, can be
predicted with significantly smaller scatter than directional
quantities like the emerging luminosity in a specific direction.

\subsection{Simulated Images and Spectra}

\begin{figure*} \ifthenelse {\boolean{blackandwhite}}{ \includegraphics[width=0.24\textwidth]{broadband_012-000-bw} \includegraphics[width=0.24\textwidth]{broadband_033-000-bw} \includegraphics[width=0.24\textwidth]{broadband_d72-000-bw} \includegraphics[width=0.24\textwidth]{broadband_040-005-bw} }{ \includegraphics[width=0.24\textwidth]{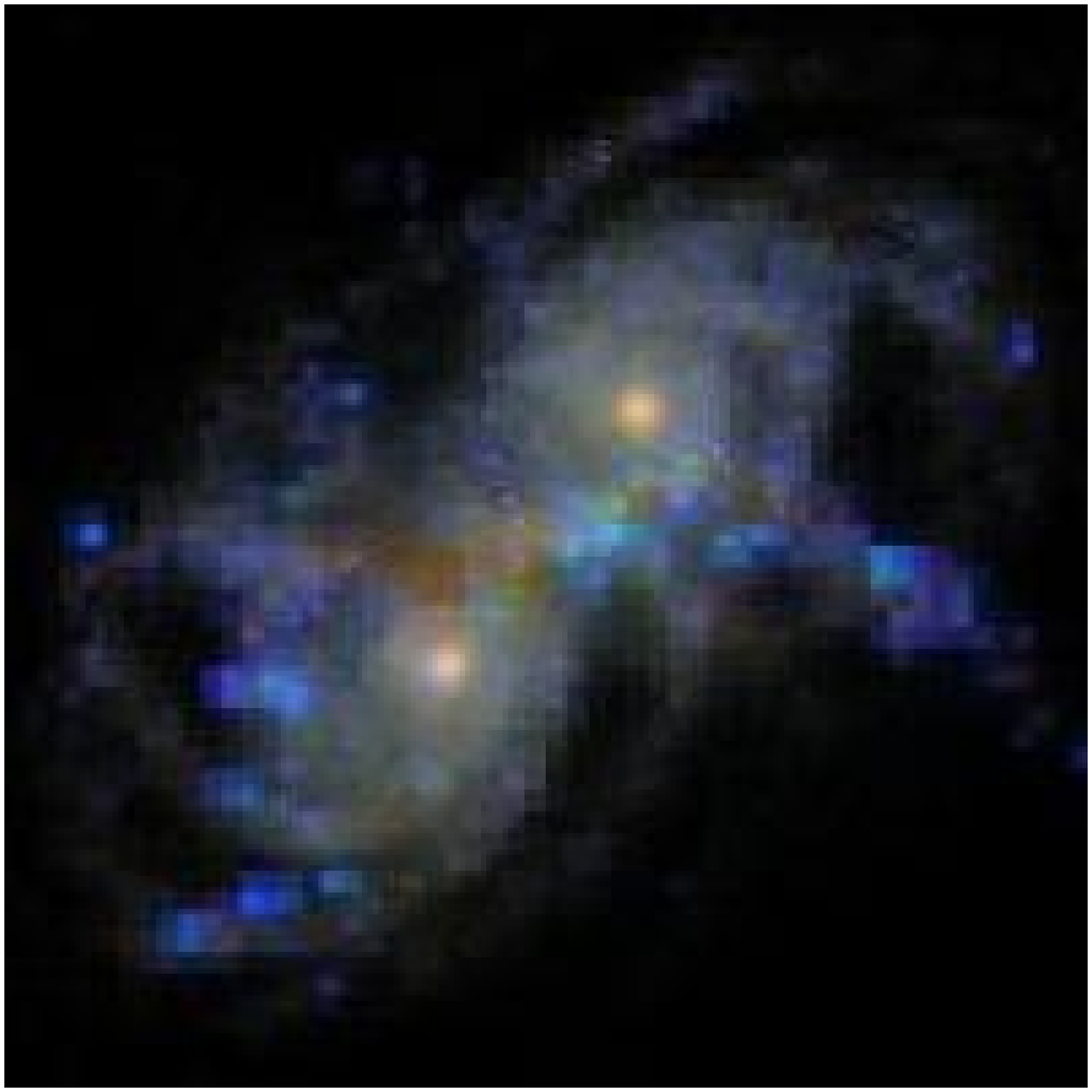} \includegraphics[width=0.24\textwidth]{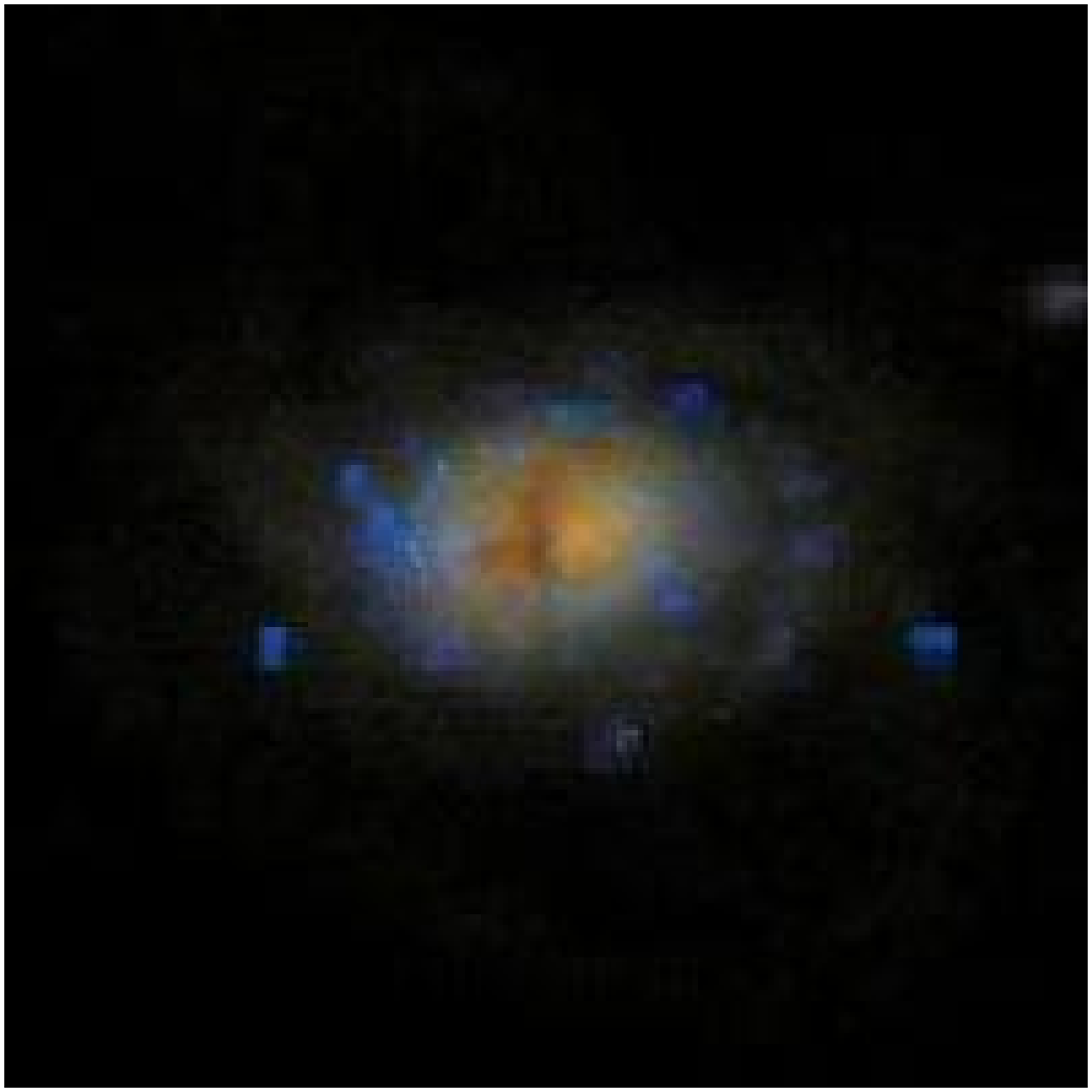} \includegraphics[width=0.24\textwidth]{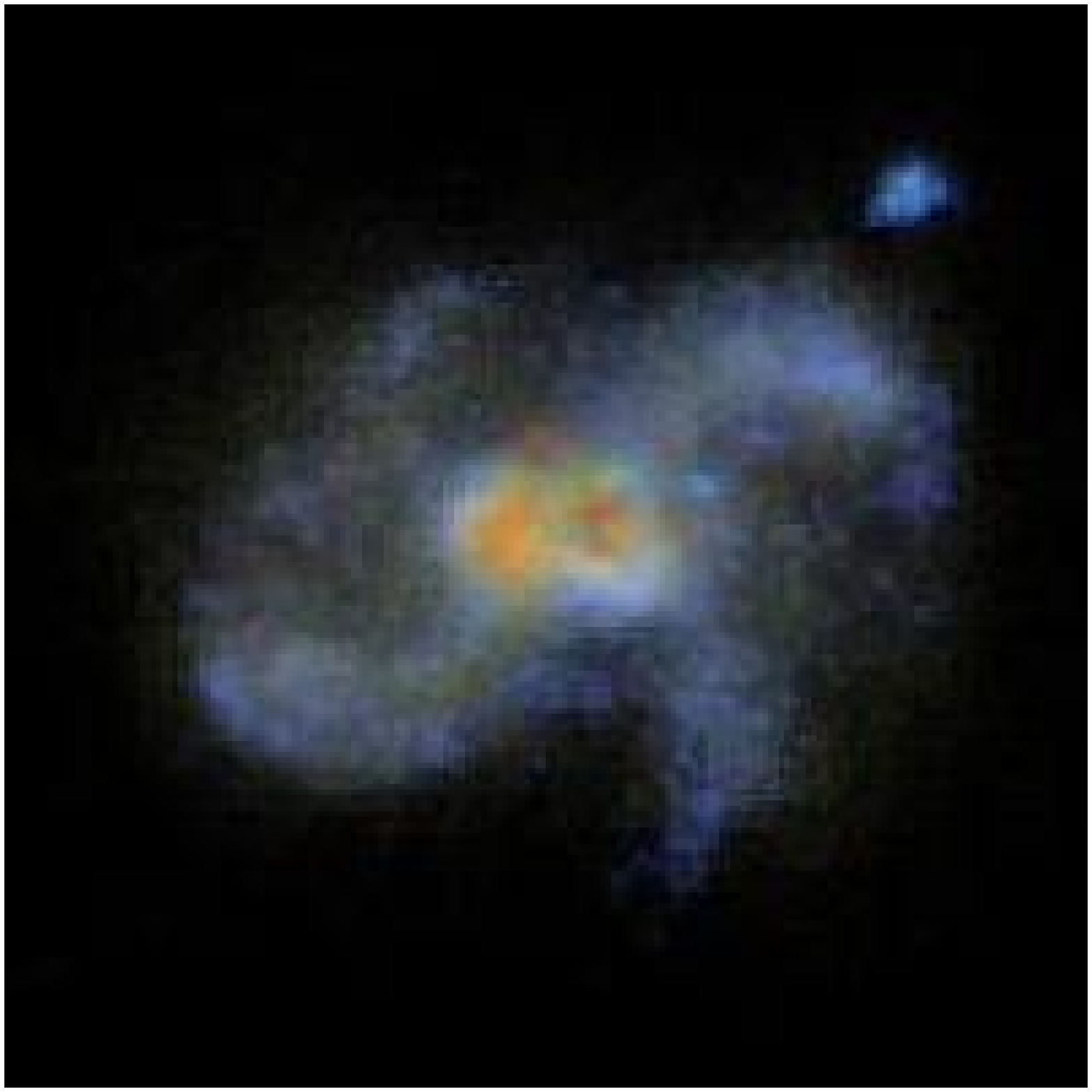} \includegraphics[width=0.24\textwidth]{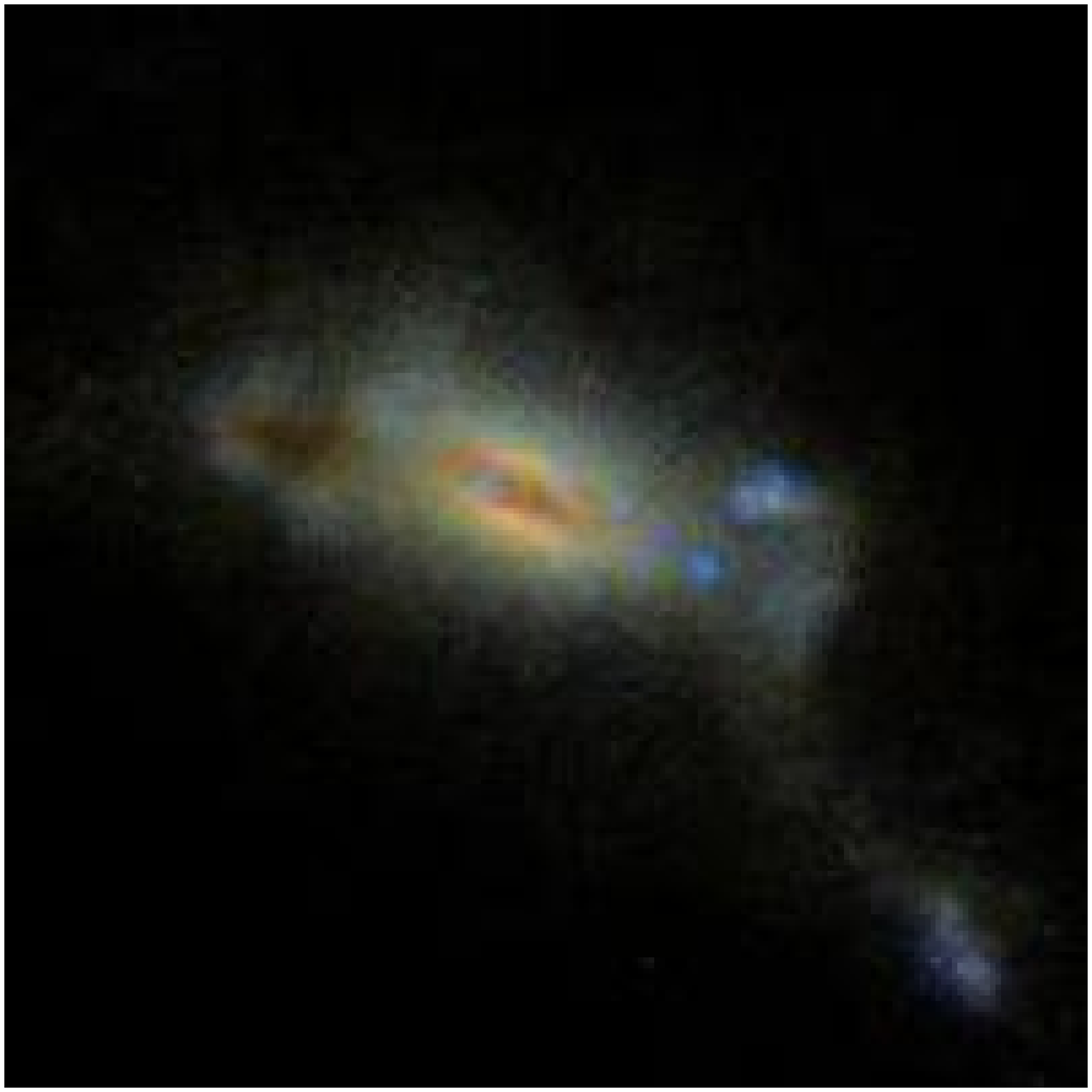} } \caption{ \label{plot_example_image} Images of the Sbc-Sbc prograde-prograde major merger at four different stages.  On the left, about $0.6\Gyr$ into the simulation, the galaxies are passing each other for the first time.  A line of induced star formation can be seen in the region between the two galaxies, where the gas in the two disks is being shocked.  The next image, at $1.6\Gyr$, shows the galaxies coming together for the second time.  This collision is almost perfectly radial, concentrating a lot of gas and dust in the center, as evidenced by the clear dust lane.  The third image, at $1.7\Gyr$, shows the final coalescence of the two galaxies, with characteristic tidal tails.  Finally, the rightmost image shows the merger remnant $2.0\Gyr$ after the start of the simulation.  While the stellar distribution is largely spheroidal, a large fraction of the gas not consumed in the starburst has now cooled and reformed a disk, visible as a dust lane. (Disk reformation was also reported by \citealt{springelhernquist04}.) The images cover $67\kpc$ squared and are in the SDSS $g$ band. (The electronic edition shows $urz$ color composite images generated using the algorithm of \citealt{lupton03}.)  } \end{figure*} \begin{figure} \ifthenelse {\boolean{blackandwhite}}{ \includegraphics[width=0.99\columnwidth]{SED_mcrx_033-00-bw}}{ \includegraphics[width=0.99\columnwidth]{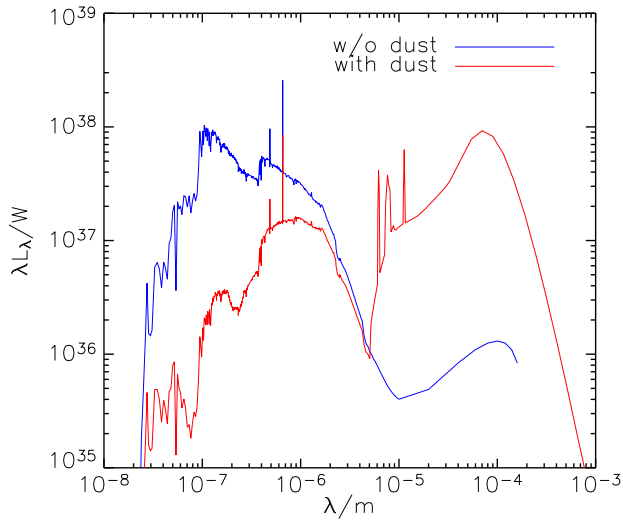}} \caption{ \label{plot_example_SED} The spectral energy distribution associated with the second image of the Sbc-Sbc prograde-prograde major merger in Figure~\ref {plot_example_image}, $1.6\Gyr$ into the simulation.  The solid (blue in the electronic edition) line shows the intrinsic stellar spectrum, while the dashed (red) line shows the emerging spectrum after taking dust attenuation into account.  At wavelengths longer than $5\micron$, the dust emission spectrum is taken from the templates of \citet{devriendtetal99}. The two emission lines in the optical are H$\alpha$ and H$\beta$. In the ultraviolet, the dust attenuation is over an order of magnitude, and the signature of the well-known ``2200\AA\ bump'' in the Milky-Way extinction curve is easily discernable.} \end{figure}

As mentioned in the model description, the outputs from the
radiative-transfer calculations are multi-wavelength images from a
number of viewpoints or, more precisely, an SED for each pixel in the
images.  To facilitate comparisons with observations, it has been
integrated into images in a number of broadband filters (listed in
Table~\ref{table-fitting-filters}) covering wavelengths from the GALEX
FUV band to the IRAC2 band on the Spitzer Space Telescope and also
collapsed into a spatially integrated SED.  In total, about 25
major-merger simulations have been run through the radiative-transfer
code, each at roughly 50 points in time.  For each point in time,
images and spectra have been generated from 11 viewpoints, equally
distributed in solid angle, and in 12 different filters. This results
in a grand total of roughly $10^5$ images and $10^4$ spectra.  The
analysis of these is the subject of a companion paper (Jonsson et al.,
in preparation), but to illustrate the outputs generated some images
are shown in Figure~\ref{plot_example_image} and a corresponding
spectrum in
Figure~\ref {plot_example_SED}.\footnote{Movies and the full set of   color images of the merger simulations can be found on the Internet   at \url{http://sunrise.familjenjonsson.org/thesis}.}

\subsection{Comparing to Observations}

\begin{figure*}   \begin {center} 
      \includegraphics[width= 0.99\columnwidth]{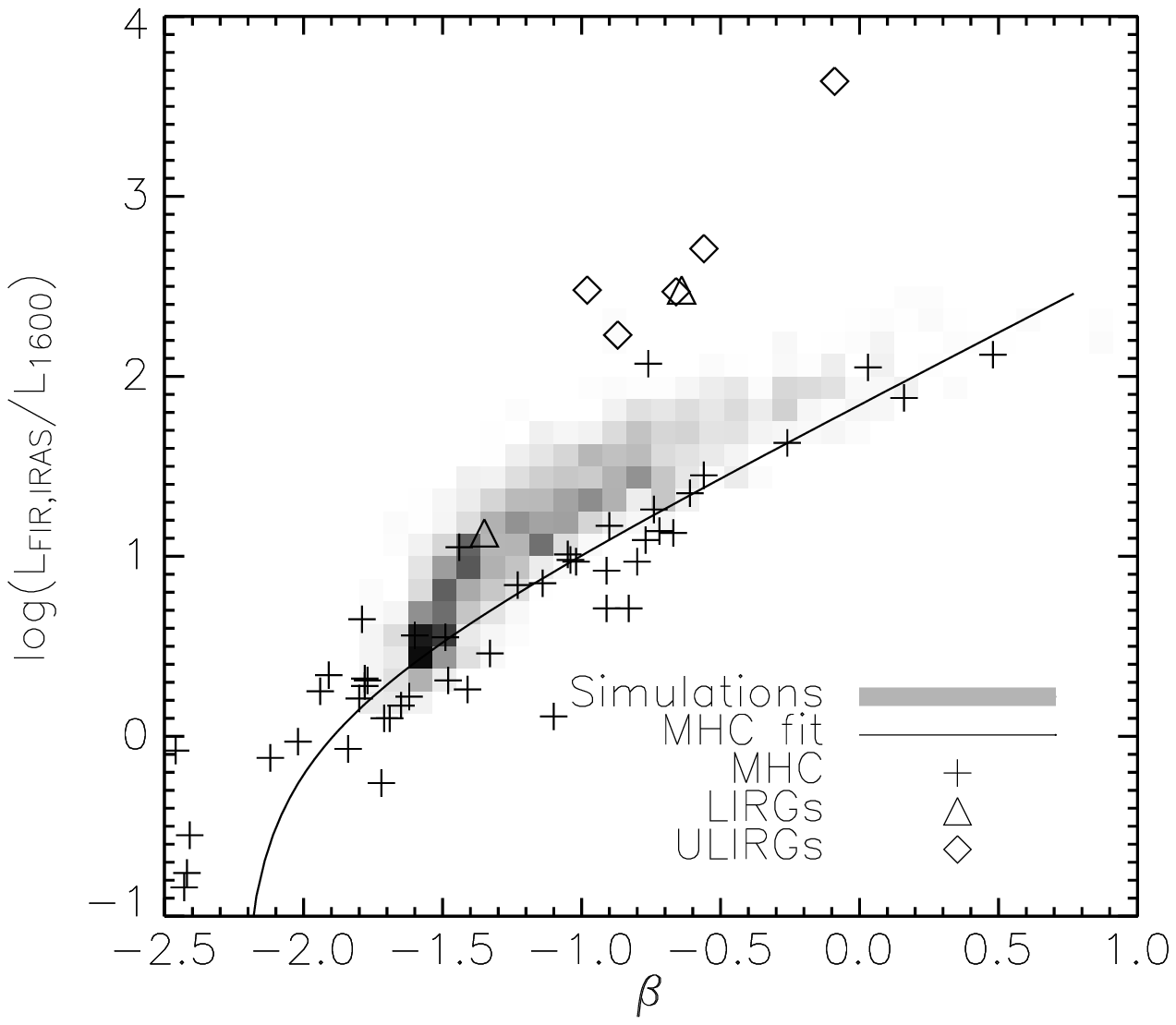}       \includegraphics[width= 0.99\columnwidth]{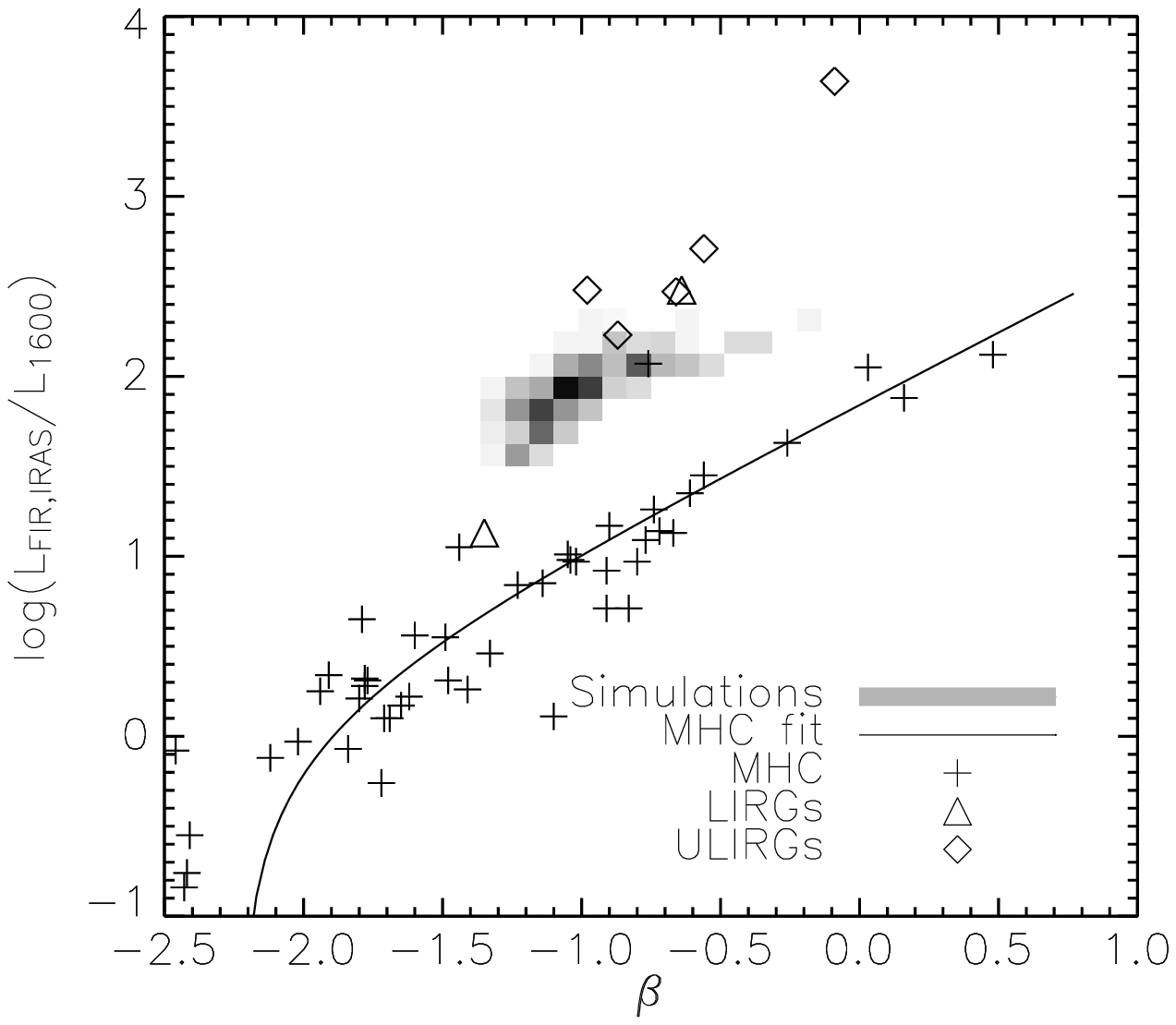} 
  \end {center}    \caption[IRX-$\beta$ correlation for different luminosities]{     \label{plot_irx_beta_luminosity}     The relation between the IR/UV flux ratio and the UV spectral     slope for the Sbc merger simulations (shaded region), compared to     the results from \citet[][hereafter MHC, crosses]{mhc99} and \citet[][     diamonds/triangles]{goldaderetal02}.  On the left, only simulated     galaxies with bolometric luminosity $L_{\mathrm{bol}} < 2\cdot     10^{11}\Lsun$ have been included. This low-luminosity sample     agrees fairly well with the MHC correlation, which is for galaxies     in this luminosity range.     On the right, only the highest-luminosity simulated galaxies, with     $L_{\mathrm{bol}} > 7\cdot 10^{11}\Lsun$ have been included.  These     points depart completely from the MHC galaxies and instead occupy the     region of LIRGs/ULIRGs from the \citet{goldaderetal02} sample.     This agreement was not a result of fitting the models, but rather     a prediction from our simulations, with no adjustment of     parameters. This and other results indicate that our simulations     provide a reasonably good replication of the properties of local     starbursts.  } \end{figure*} 

In order to determine how well our simulations mimic actual starburst
galaxies, they must be compared to observations.  While this comparison
is described in \citet{pjthesis-nourl} and is the subject of a
companion paper, example results are included here to illustrate that
the simulations appear to replicate the properties of observed
galaxies.  After accounting for dust effects, the simulations have an
absolute $r$-band magnitude in the range -21.5 to -22.5 and a $u - r$
color of 1.3 to 2.2, falling in the region of bright, blue galaxies in
the Sloan Digital Sky Survey \citep{baldryetal04}.  Even $1 \Gyr$ after
the merger, there is enough ongoing star formation in the merger
remnant that it is among the blue galaxies.  In recent simulations,
\citet{springeletal04agnb, springeletal04} have shown that feedback
from an active galactic nucleus can help truncate star formation, which
would let the merger remnant redden more quickly.  In any case, the
inclusion of dust is crucial for the agreement; without it, the systems
are far too bright and blue to agree with observed galaxies.

The simulations were also compared to observations by 
\citet[][hereafter MHC]{mhc99}
and \citet{h98}.  Both of these studies looked at correlations between
dust attenuation and other properties of starburst galaxies.  MHC
looked at the relation between dust absorption, indicated by the
far-infrared over ultraviolet flux ratio, and the ultraviolet spectral
slope.  In a sample of moderately luminous starburst galaxies observed
by International Ultraviolet Explorer (IUE) and IRAS, they found a
fairly tight correlation between the two parameters.  However,
\citet{goldaderetal02} examined a small sample of more luminous LIRGs
and ULIRGs, and found that they depart from the relation seen by MHC in
the sense that their UV color is too blue for their infrared
luminosity.  When the same quantities are extracted from the
simulations, the agreement, shown in
Figure~\ref{plot_irx_beta_luminosity}, is remarkable.  The simulations
follow the MHC relation when similar-luminosity systems are selected. 
In contrast, when the highest-luminosity subsample of the simulated
galaxies is selected, they are found in the same region as the
LIRGs/ULIRGs of \citet{goldaderetal02}. It should be emphasized that
this is not a result of fitting the simulations to these observations,
but rather a prediction of our initial conditions which were selected,
a priori, to be realistic for local spiral galaxies. Unlike in earlier
studies \citep{gcw97}, this agreement is contingent on the use of
Milky-Way dust.  SMC-type dust leads to far too red ultraviolet slopes
and is inconsistent with the MHC relation in our simulations.

\citet{h98} studied correlations between dust absorption, infrared and
ultraviolet luminosity, metallicity, absolute magnitude, etc.  The
agreement with those results, though not shown here, is also
encouraging.  These, while not exhaustive tests, indicate that our
simulations provide a good replication of local starbursts.  A future
paper (Jonsson et al., in preparation) will explore the comparison
between simulations and observations in detail.

The simulations can also be used to study the performance of
star-formation indicators in the presence of dust.  While the
far-infrared luminosity provides a robust indication of the
star-formation rate for starburst systems, both the $\halpha$ and the
ultraviolet luminosity suffer from severe dust effects
\citep{pjthesis-nourl}.  Published dust corrections based on the Balmer
line ratio \citep{calzettietal94} or the ultraviolet spectral slope
\citep{bellkennicutt01} improve the estimate, but the star-formation
rate can still be underestimated by up to an order of magnitude at the
highest luminosities, where the dust corrections perform the worst. A
detailed analysis of the star-formation indicators in the presence of
dust is planned for a future paper.

\subsection{Dust Attenuation}

\begin{figure} 
  \ifthenelse {\boolean{blackandwhite}}{     \includegraphics[width=0.99\columnwidth]{l-att-bw}}{     \includegraphics[width=0.99\columnwidth]{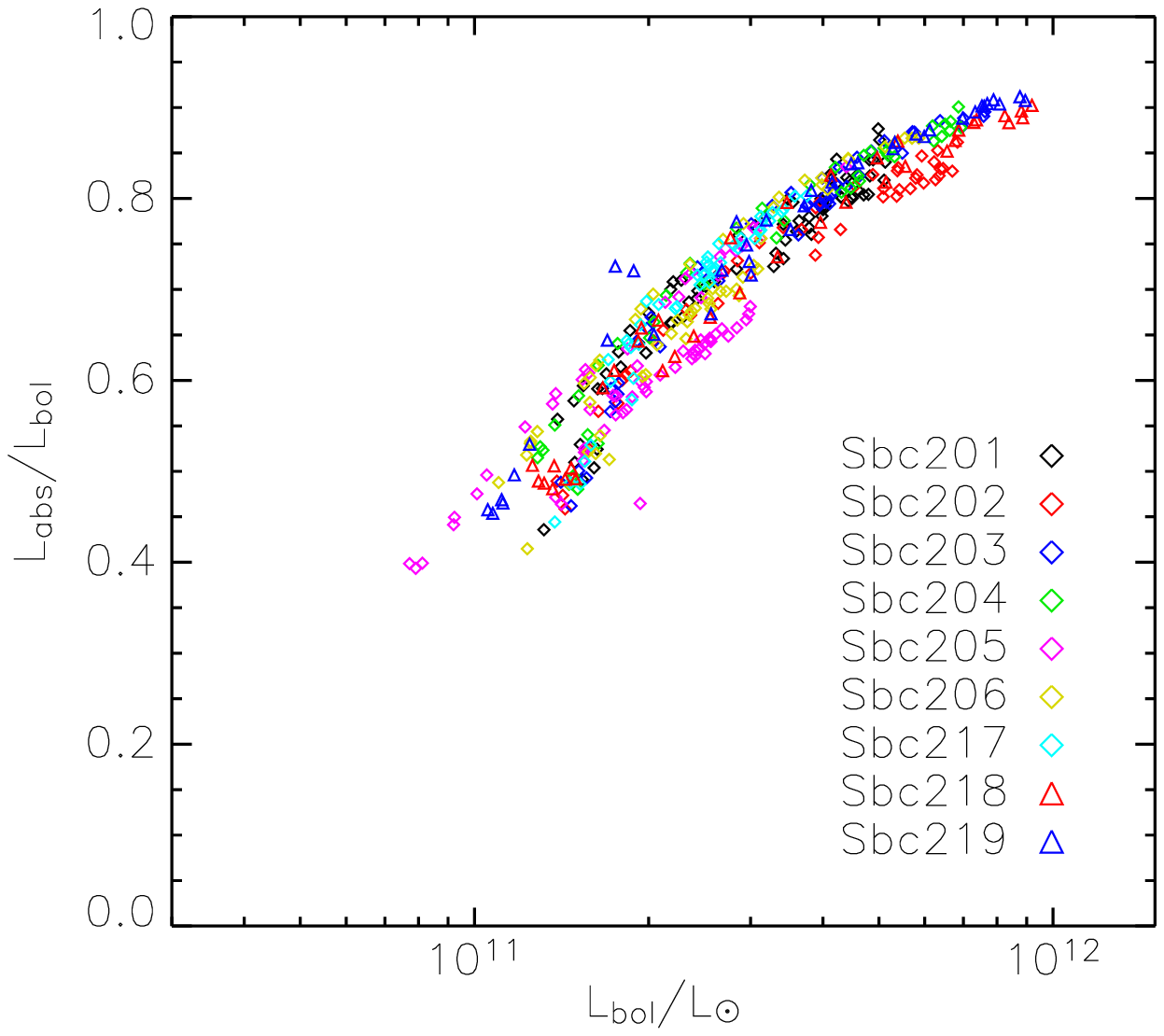}} \caption[Relation between bolometric luminosity and attenuation] { \label{plot_luminosity_attenuation} The relation between bolometric luminosity and attenuation (fraction of bolometric luminosity absorbed by dust, averaged over all directions) for the nine different Sbc-Sbc major merger simulations on various orbits.  Each point is a simulation snapshot.  For a given luminosity, the attenuations lie in a narrow range, regardless of encounter geometry, stage of merger and the fact that different simulations cover different luminosity intervals.  All simulations were started with solar metallicity gas.  (In the electronic edition, the different colors identify the different simulations in Appendix A of \citealp{tjthesis-nourl}.)  } \end{figure}

Looking at the nine Sbc simulations shown in
Figure~\ref{plot_luminosity_attenuation}, there is a tight correlation
between the bolometric luminosity of the system and the bolometric dust
attenuation, averaged over all directions.  For a given bolometric
luminosity a uniquely determined fraction of luminosity is absorbed by
the dust; it does not seem to matter if the system consists of two
barely-interacting disks, a merger-driven starburst (on various
orbits), or a post-starburst remnant.  Simulations using galaxies of
different mass or metallicity follow similar  correlations offset to
higher or lower attenuation.

Theoretically, it is expected that luminosity (for starbursting systems
largely determined by star-formation rate) and dust absorption should
correlate in the simulations, as both of these quantities are driven by
gas density. Concentrating the gas to larger densities will increase
the star-formation rate, and hence the bolometric luminosity, through
the Schmidt law used.  Concentrating the gas also increases the gas,
and hence dust, column density, increasing dust absorption. Several
observational studies have also concluded that dust attenuation
generally seems to increase with galaxy luminosity and star-formation
rate \citep{wangheckman96, adelbergersteidel00, hopkinsetal01,
sullivanetal01, vijhetal03, buatetal04}. In the following section, a
toy model for estimating dust absorption will be presented. This will,
in turn, motivate the fitting formula used to estimate the dust
absorption in the simulations.

\subsection{A Toy Model for Dust Attenuation}

In order to come up with a simple model for how dust absorption should
depend on luminosity, mass, and metallicity of a galaxy, consider a
constant-density sphere of star-forming gas. For a sphere, the density
is given by 
\begin{equation}
\label{equation-density} \rho \propto M_g R^{ - 3 } \> ,
\end{equation}
 where $M_g$ is the gas mass and $R$ is the radius. According to the
Schmidt law used to estimate the star-formation rate in the
simulations, the star-formation rate density is 
\begin{equation}
\dot \rho_\star \propto \rho^{ 3 / 2 } \>.
\end{equation}
 The total star-formation rate $\dot M_\star$ is thus 
\begin{equation}
\label{equation-sfr} \dot M_\star \propto M_g^{ 3 / 2 } R^{ - 3 / 2 }
\>.
\end{equation}
 The optical depth of dust in the sphere will depend on the column
density and the metallicity $Z$ of the gas, 
\begin{equation}
\label{equation-tau} \tau \propto Z R \rho \propto Z M_g R^{ - 2 } \>.
\end{equation}
 Eliminating $R$ using equation~\ref{equation-sfr},  we get 
\begin{equation}
\label{equation-tau-final} \tau \propto Z \dot M_\star^{ 4 / 3 } M_g^{
- 1 } \propto Z L^{ 4 / 3 } M_g^{ - 1 } \> ,
\end{equation}
 where the last proportionality comes from assuming that the bolometric
luminosity $L$ is proportional to the star-formation rate. Finally,
once the optical depth is determined, the absorbed fraction of
luminosity (i.e., the attenuation) in a medium where luminous and
absorbing material is uniformly mixed is given by
\citep[e.g.][]{calzettietal94}
\begin{equation}
\label{equation-toymodel} { { L_{ \rm { abs } } } \over L } = 1 - { { 1
} \over { \tau } } \left ( 1 - e^{ - \tau } \right ).
\end{equation}
 Actually, equation~\ref{equation-toymodel} is appropriate for a
plane-parallel slab, not a sphere.  However, the purpose of the toy
model is to find a simple, physically motivated, fitting formula.  As
will be shown in the next section, equation~\ref{equation-toymodel}
describes the behavior of the simulations well. Hence, in the interest
of simplicity, it will be used.

This toy model has obvious limitations: It neglects scattering, and the
exponents $4 / 3$ and $- 1$ depend on the assumed geometry.
Furthermore, the model is really more representative of an individual
star-forming region than an entire galaxy, so included in the constant
of proportionality in equation~\ref{equation-tau-final} is the number
of such regions in the galaxy. If this number depends on the properties
of the galaxy, a reasonable assumption, it will change the  dependence
on the different quantities in equation~\ref{equation-tau-final}. 
Finally, the attenuation of the bolometric luminosity is an average of
the attenuation at all wavelengths, and because the system in general
is optically thick at short wavelengths and more or less optically thin
at longer wavelengths, the behavior is more complicated than the simple
equation~\ref{equation-toymodel}.   Real galaxies are thus more
complicated than our assumptions, but our simple model gives a general
description of the trends for the effects of dust.

\subsection{Fitting Functions}

While $L \propto \dot M_\star$ is a good approximation for starbursting
galaxies, where young stellar populations dominate the luminosity, this
assumption is not good in general.  For this reason, simultaneous
dependence on both $L$ and $\dot M_\star$ will be retained.

Guided by the toy model, one would expect the \emph{gas} mass to be the
dominant factor in the relation.  However, fits using total
\emph{baryonic} (gas plus stars) mass, $M_b$, instead of gas mass had
lower residuals.  For this reason, baryonic mass is used for the fits.

Motivated by the above toy model and these considerations, the general
fitting function to be used for the simulations is 
\begin{equation}
\label{equation-fitting} \begin{array} { l } { { L_{ \rm { abs } } }
\over L } = 1 - { { 1 } \over { \tau } } \left ( 1 - e^{ - \tau }
\right ) \> \mbox { , where } \\ \tau = \tau_0 \left ( { Z \over 0.02 }
\right )^{ \! \alpha } \! \left ( { L \over { 10^{ 11 } \Lsun } }
\right )^\beta \! \left ( { { \dot M_\star } \over { \Msun \yr^{-1} } }
\right )^{ \! \gamma } \! \! \left ( { { M_b } \over { 10^{ 11 } \Msun
} } \right )^\delta \> , \end{array}
\end{equation}
 and the Greek letters denote free parameters.

\begin{deluxetable}{ccccccl} 
\tablecolumns{7}   \tablecaption{Fits to the bolometric attenuation.     \label{table-fitting-bolometric}}   \tablehead{     \colhead{$\tau_0$\tablenotemark{a}} &     \colhead{$\alpha$\tablenotemark{a}} &     \colhead{$\beta$\tablenotemark{a}} &     \colhead{$\gamma$\tablenotemark{a}} &     \colhead{$\delta$\tablenotemark{a}} &     \colhead{$\sigma$\tablenotemark{b}} &     \colhead{Comment}} \startdata 1.25 & 1.02 & 0.41 & 0.39 & -0.82 & 0.04 & Full fit \\ 0.93 & 1.10 &       & 0.61 & -0.68 & 0.04 & Indep. of $L$ \\ 2.10 & 0.91 & 1.10 &       & -1.03  & 0.05 & Indep. of $\dot M_\star$ \\ 2.09 & 0.94 &      &      &       & 0.14 & Dep. on $Z$ only.\\ 2.34 &      & 0.28 &      &       & 0.12 & Dep. on $L$ only.\\ 1.70 &      &      & 0.26 &       & 0.10 & Dep. on $\dot M_\star$     only. \\ 2.21 &      &      &      & \phs0.16  & 0.16 & Dep. on $M_b$ only. \\ 0.32 & 0.30 & -0.15 & 0.82 & -0.52 & 0.05 & Full fit with $M_g$\tablenotemark{c} \enddata \tablenotetext{1}{ Parameter in equation~\ref{equation-fitting}.} \tablenotetext{2}{ The standard deviation of the scatter around the fit.} \tablenotetext{3}{ This fit was performed with gas mass, instead of     baryonic mass, driving the parameter $\delta$ in      equation~\ref{equation-fitting}.} \tablecomments{The attenuation is the fraction of bolometric     luminosity which is absorbed, averaged over all directions.} \end{deluxetable}

The fits were done as simple $\chi^2$ minimizations, with one point for
each simulation snapshot in the simulations listed in
Table~\ref{table_mass_metallicity}.  As there were nine different Sbc
merger simulations and only one of each of the simulations using the
other galaxy models, the Sbc simulations were given only $1 / 9$ the
weight in the fits.  This was done to avoid giving undue weight to the
massive, bright, and gas-rich Sbc mergers. Because the time between
saved simulation snapshots was varied in order to capture short-lived
stages, each simulation snapshot was also weighted in proportion to its
``time of influence'', i.e. the time to preceding and following
snapshots.  Apart from this weighting, constant errors on the points
were assumed.  Table~\ref{table-fitting-bolometric} shows the
parameters resulting from fits of the bolometric attenuation to the
simulations under different constraints.  The quantities come in with
different powers from those in equation~\ref{equation-toymodel}, but,
as noted earlier, this is not surprising.

\begin{figure} 
    \includegraphics[width=0.99\columnwidth]{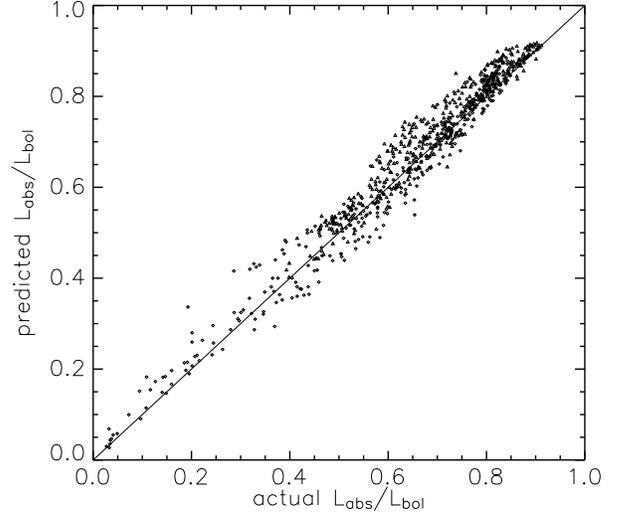}
\caption{ \label{plot_lattres-full} Actual attenuation (fraction of bolometric luminosity absorbed by dust) in the simulations, compared to the attenuation predicted by the full fit of equation~\ref{equation-fitting} (the first line of Table~\ref{table-fitting-bolometric}).  Each point is a simulation snapshot. 
} \end{figure}

A plot of the actual attenuations against those predicted from the full
fit is shown in Figure~\ref{plot_lattres-full}. The fit describes the
behaviour of the simulations well, with a $1 \sigma$ scatter of about
0.04 in the absorbed fraction.  Since this fit describes the fraction
of bolometric luminosity absorbed by dust, it can be used to predict
the dust luminosity of these systems.  It should be emphasized that the
quantities used are those of the system, implying that when the merging
galaxies are still distinct it is the aggregate luminosity, mass, and
star-formation rate of the two which is used.

Table~\ref{table-fitting-bolometric} contains additional parameter sets
besides the complete fit. These are appropriate if some of the
quantities which go into equation~\ref{equation-fitting} are unknown.
With a modest increase in the scatter around the relation, one of the
largely complimentary quantities $L$ or $\dot M_\star$ can be excluded.
In these cases, the power of the quantity not excluded increases to
assume the role of the excluded quantity.  With significantly increased
scatter, the attenuation can also be predicted from one quantity only. 
This is presumably because metallicity, luminosity, and mass are
intrinsically correlated in galaxies.  It is difficult to say to what
degree these single-parameter correlations are affected by our small
set of galaxy models.

Finally, the last line of Table~\ref{table-fitting-bolometric} contains
the parameters obtained when the fit was done using the gas mass.  The
fit has about 50\% larger scatter than when  the baryonic mass was
used, so it still provides a useful description of the data.  However,
the dependence on the quantities is strange, $\beta$ is negative which
means that more luminous galaxies should have \emph{smaller} dust
attenuation.  It is as if luminosity has assumed part of the function
of the baryonic mass.

\subsection{Fits At Specific Wavelengths \label{fit-filters}}

\begin{deluxetable}{cccccccl} 
  \tablecolumns{7}   \tablecaption{Fits to the dust attenuation in specific filters.     \label{table-fitting-filters}}   \tablehead{     \colhead{$\tau_0$\tablenotemark{a}} &     \colhead{$\alpha$\tablenotemark{a}} &     \colhead{$\beta$\tablenotemark{a}} &     \colhead{$\gamma$\tablenotemark{a}} &     \colhead{$\delta$\tablenotemark{a}} &     \colhead{$\sigma$\tablenotemark{b}} &     \colhead{$\sigma_i$\tablenotemark{c}} &     \colhead{Filter}} \startdata 
\phn7.67 & 3.33 &    -0.04 & 0.29 & -0.74 & 0.07 & 0.04 & GALEX FUV \\ \phn5.03 & 2.45 & \phs0.03 & 0.36 & -0.61 & 0.07 & 0.04 & GALEX NUV \\ \phn1.84 & 1.52 & \phs0.45 & 0.28 & -0.70 & 0.10 & 0.05 & SDSS $u$ \\ \phn1.27 & 1.39 & \phs0.55 & 0.19 & -0.70 & 0.11 & 0.05 & SDSS $g$ \\ \phn0.94 & 1.36 & \phs0.61 & 0.16 & -0.71 & 0.12 & 0.05 & SDSS $r $ \\ \phn0.78 & 1.34 & \phs0.65 & 0.12 & -0.70 & 0.11 & 0.05 & SDSS $i $ \\ \phn0.68 & 1.37 & \phs0.69 & 0.10 & -0.72 & 0.11 & 0.05 & SDSS $z $ \\ \phn0.53 & 1.43 & \phs0.72 & 0.09 & -0.74 & 0.10 & 0.05 & 2MASS J \\ \phn0.42 & 1.54 & \phs0.76 & 0.09 & -0.78 & 0.09 & 0.05 & 2MASS H \\ \phn0.32 & 1.62 & \phs0.77 & 0.10 & -0.80 & 0.08 & 0.05 & 2MASS Ks \\ \phn0.19 & 1.78 & \phs0.74 & 0.14 & -0.82 & 0.06 & 0.04 & Spitzer IRAC1\\ \phn0.14 & 1.95 & \phs0.71 & 0.18 & -0.86 & 0.06 & 0.04 & Spitzer IRAC2\\ 23.9\phn &      & \phs0.70 &      &      & 0.12 & 0.10 & 1900\AA (WH96)\tablenotemark{d} \\ \phn4.0\phn &   & \phs0.49 &      &      & 0.15 & 0.11 & B (WH96)\tablenotemark{d} \enddata \tablenotetext{1}{ Parameter in equation~\ref{equation-fitting}.} \tablenotetext{2}{ The standard deviation of the total scatter around the   fit, including the variation with viewing angle.} \tablenotetext{3}{ The standard deviation of the intrinsic scatter   around the fit, i.e. excluding the variation with viewing angle.} \tablenotetext{4}{ Fits done as  a comparison to the study by \citet{wangheckman96}, described in section~\ref {section_observations}.} \tablecomments{Fits were performed using all parameters. The scatter in the far-ultraviolet bands is suppressed because most attenuations are close to 1. 
} \end{deluxetable}

\begin{figure} \includegraphics[width=0.99\columnwidth]{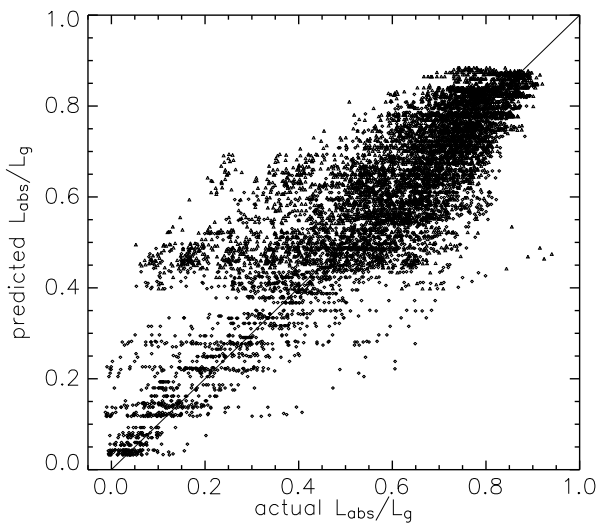} \caption{ \label{plot_lattres-g} Actual attenuation in the simulations versus the attenuation predicted by equation~\ref{equation-fitting}, but now for the luminosity in the Sloan Digitized Sky Survey $g$ band.  Unlike Figure~\ref{plot_lattres-full}, which shows absorbed energy averaged over all lines of sight, this figure shows the line-of-sight attenuation. Each point is a simulation snapshot from a certain line of sight.  The slightly negative attenuations result from preferential scattering out of the plane of the disk in the initial galaxies. This fit has significantly larger scatter than the fit to bolometric attenuation, a result mainly because of the large variation of attenuation with viewing angle, but provides a good description of the behavior averaged over all directions. } \end{figure}

The fits in the previous section are useful for predicting the infrared
dust luminosity of the systems, but they cannot be used to predict the
attenuation of radiation emerging from the system observed at a
specific wavelength.  Furthermore, in order to be able to make
predictions of the luminosity inferred by an observer, the variation in
attenuation with line of sight must be considered. 
Table~\ref{table-fitting-filters} contains fits, also using
equation~\ref{equation-fitting}, to the attenuation in each of the
bandpasses included in the calculation.  Here, each point in time of a
simulation is associated with 11 different attenuations, one for each
of the different viewing angles calculated.  These fits have
significantly larger scatter than the fit to the bolometric attenuation
in Table~\ref{table-fitting-bolometric}, but this is largely due to the
variation of attenuation over different lines of sight.  When the
attenuation is averaged over all lines of sight, the fit is unchanged,
but the scatter is much smaller, only marginally larger than for the
fit to bolometric attenuation.  An example of a fit, for the SDSS $g$
band, is shown in Figure~\ref{plot_lattres-g}.

\subsection{Observations of Dust Attenuation
\label{section_observations}}

Several previous studies have attempted to estimate the dust
attenuation in observed galaxies nature, and our simulations can be
compared to these results.

\begin{figure*}   \ifthenelse {\boolean{blackandwhite}}{ \includegraphics[width=0.99\columnwidth]{lattres-1900-bw} \includegraphics[width=0.99\columnwidth]{latt-b-bw}}{ \includegraphics[width=0.99\columnwidth]{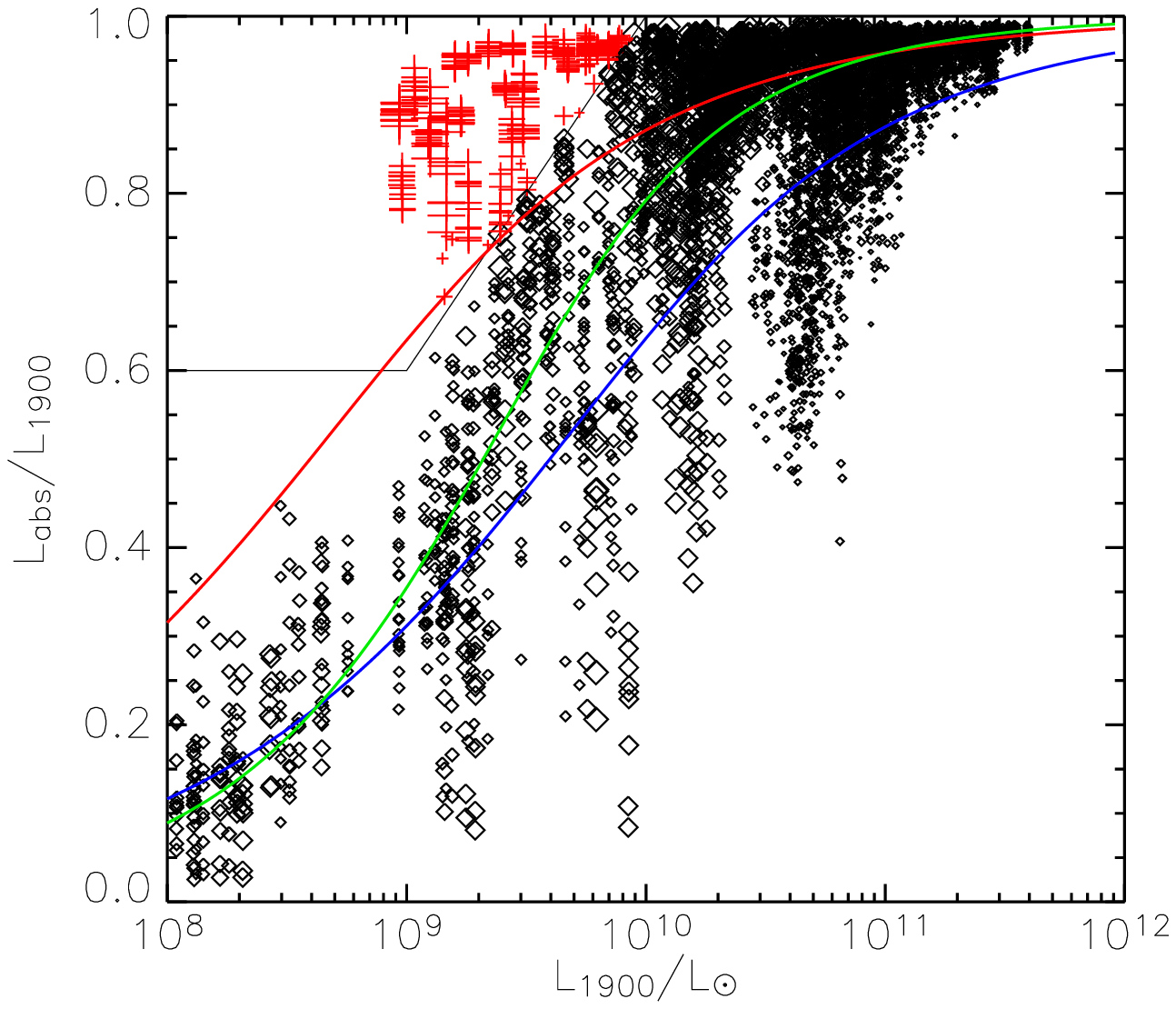} \includegraphics[width=0.99\columnwidth]{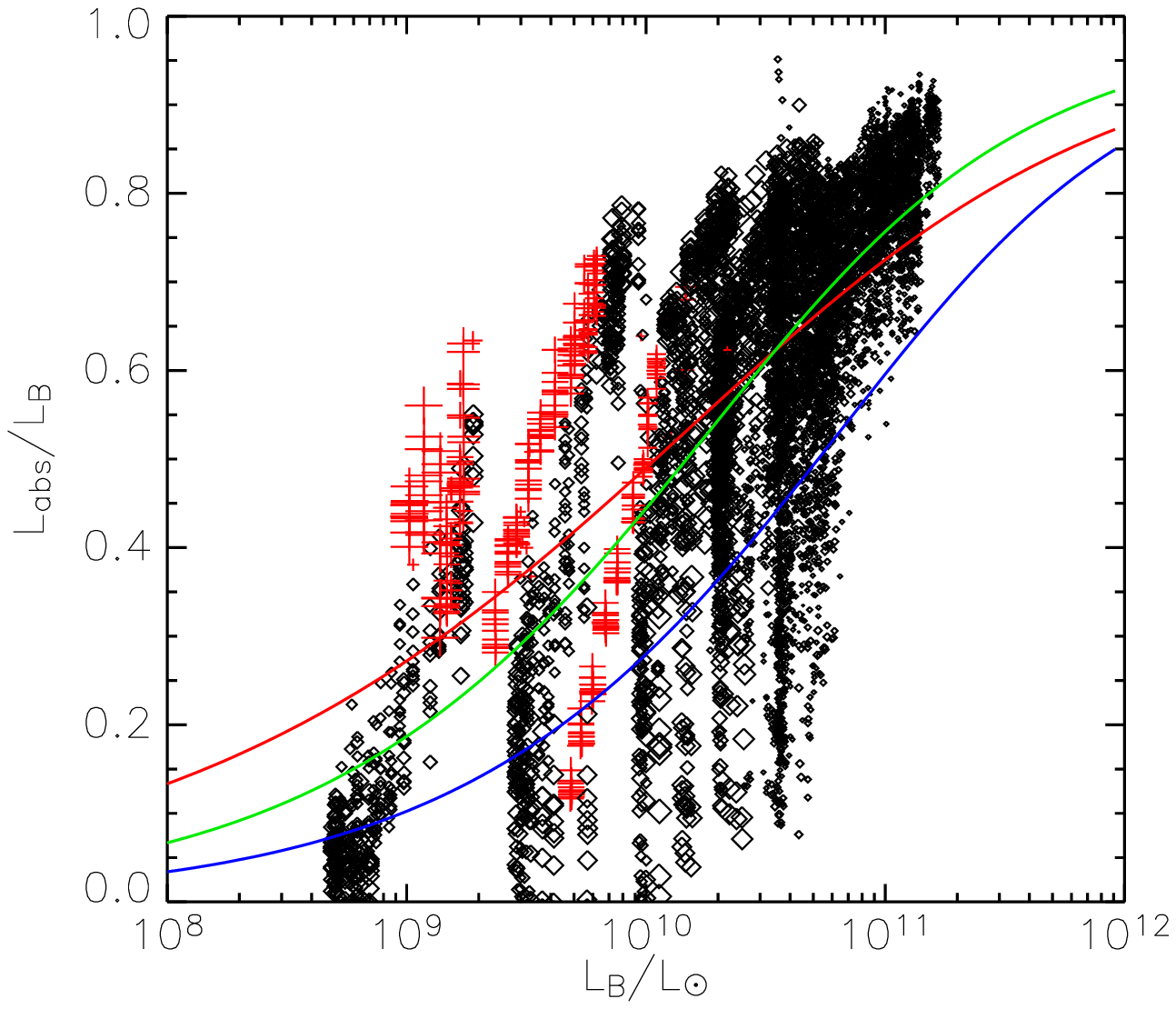}} \caption{ \label{plot_lattres-1900} \label{plot_lattres-b} Comparison with \citet {wangheckman96}.  On the left, luminosity at $1900\Angstrom$ versus attenuation at $1900\Angstrom$ for the simulations, on the right luminosity in the B band versus attenuation in the B band.  The size of the symbols are proportional to the fitting weight of the points. The dashed (green in the Electronic Edition) lines show the fits marked ``WH96'' in Table~\ref{table-fitting-filters}.  The points in the upper left region in the left plot, marked by the solid line, consist of merger remnants and were excluded from the fit for reasons explained in Section~\ref {fit-filters}. Because these points are not nicely delineated in the right plot, they are also marked as (red) crosses. (The dotted (red) lines show the fits obtained if these points are included in the fit.)  The dot-dashed (blue) lines show the WH96 fits. In the ultraviolet, the WH96 fit agrees fairly well with the simulated galaxies for luminosities $< 10^{10}\Lsun$, the maximum UV luminosity of the WH96 galaxies, but underpredicts the attenuation at high luminosities.  In the B band, the WH96 fit has a luminosity dependence similar to our fit, but is normalized lower.  } \end{figure*}

\citet{wangheckman96} examined the dust content in late-type galaxies
based on their UV/FIR flux ratio and fit the derived optical depth to a
power-law dependence on luminosity.  To compare to their study, a fit
of the ultraviolet attenuation using only a dependence on ultraviolet
luminosity is also performed. This fit is shown in
Figure~\ref{plot_lattres-1900}.  This fit was complicated by the fact
that when using only luminosity instead of all parameters to predict
the dust attenuation, there is a significant difference between the
initial merging galaxies and the merger remnants. Especially in the
smaller G2, G1, and G0 mergers, the merger remnants have significantly
higher dust attenuation for their luminosity compared to the earlier
stages.  Furthermore, because the simulated galaxies spend a lot of
time as merger remnants (limited by how long the simulation has been
run), this stage has significant weight when fitting.  These points can
be seen in the upper left of Figure~\ref{plot_lattres-1900}. As
discussed below, our model likely overestimates the dust attenuation in
the merger remnants.  For this reason, and also because WH96 did not
include early-type galaxies, these low-luminosity, high-attenuation
points were excluded in this fit.  (If these points are included, the
resulting fit, also shown in Figure~\ref{plot_lattres-1900}, does not
provide a good description for the low-luminosity pre-merger and
merging galaxies.)

After excluding these points, the fit yields $\beta = 0.70$, consistent
with the result of WH96 who obtained $\beta = 0.5 \pm 0.2$. In terms of
the normalization, our fit results in $\tau_0 = 23.9$, at a luminosity
of $10^{ 11 } \Lsun$. Rescaled to the WH96 luminosity zeropoint of $4.5
\times 10^9 \Lsun$, it corresponds to a UV optical depth of 2.7, while
their result was $1.7 \pm 0.6$. The simulated galaxies seem to have
larger optical depth for the same luminosity, possibly reflecting the
fact that they not really  ``normal'' spirals. However, the WH96
galaxies  were essentially limited to $L_{ 1900 } < 10^{ 10 } \Lsun$,
and in this luminosity range, the WH96 fit is a good description also
of the simulated galaxies.  The intrinsic scatter in this fit is much
larger than when all the parameters are used.

WH96 also presented their results in terms of the B-band luminosity
and, correspondingly, the attenuation in the B band.  To compare to
these result, a fit to the simulations was done in the B band (using
the prescription of \citealt{fukugitaetal96} to transform the SDSS
magnitudes into a B magnitude). The same points were excluded for this
fit, also shown in Figure~\ref{plot_lattres-b}, as for the ultraviolet
fit.  This results in $\beta = 0.49$, in excellent agreement with WH96.
 Their normalization is lower also in this case.  Our result was
$\tau_0 = 4.0$, which, rescaled to the WH96 luminosity of $1.3 \times
10^{ 10 } \Lsun$, corresponds to an optical depth of 1.5. Their result
was $0.8 \pm 0.3$.

A similar study was performed by \citet{vijhetal03}, who estimated the
UV attenuation in a sample of Lyman-break galaxies using
radiative-transfer models of clumpy dust shells.  They basically
obtained  
\begin{equation}
1 - { { L_{ \rm { abs } } } \over { L } } \propto L^{ - 0.95 \pm 0.5 }
\> ,
\end{equation}
  at $1600 \Angstrom$. In optically thick situations, this would
correspond to $\beta = 0.95 \pm 0.5$ in our formulation, also
consistent with our result at $1900 \Angstrom$.

\begin{figure} \includegraphics[width=0.99\columnwidth]{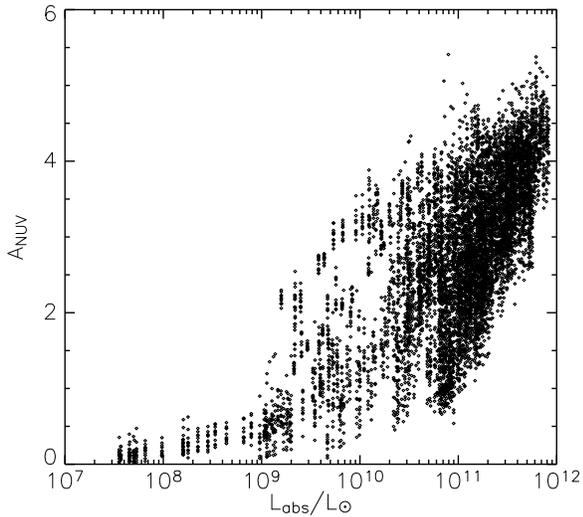} \caption{ \label{plot_latt-nuv} Luminosity absorbed by dust versus attenuation in the GALEX NUV band (in magnitudes) for the simulated galaxies. This plot can be directly compared to Figure~3 of \citet {buatetal04}.  Their infrared-selected galaxy sample shows a very similar distribution, while their UV-selected sample has lower attenuations.  Given that our simulated galaxies are bright starbursts, the infrared selection is likely more appropriate, at least at higher luminosities. For $L_\mathrm{abs}<10^{10}\Lsun$, the \citet{buatetal04} IR-selected sample contains very few galaxies, probably a selection effect, but the ones present lie in the same region as the simulations.  } \end{figure}

\citet{buatetal04} compared the dust attenuation, in the GALEX NUV
band, of two samples selected in the near-ultraviolet and far-infrared.
 While they saw an increase in dust attenuation with luminosity in both
samples, they did not attempt to determine the luminosity dependence. 
No attempt has been made to replicate their selection criteria in our
simulated galaxies either, but the distribution of attenuation with
luminosity in the simulations (shown in Figure~\ref{plot_latt-nuv}) is
similar to that of their sample.

\subsection{Model Limitations}

While our suite of simulations is the most ambitious effort to
self-consistently model dust extinction in starbursts so far, our model
has limitations.  For one, dust is known to be clumpy or patchy on
small scales, something which is not taken into account in the
simulations.  Previous studies \citep{wittgordon96,wittgordon00, gcw97}
have indicated that clumping has important implications for the effects
of dust.  While the large-scale structure of the dust distribution in
the galaxies is determined by the hydrodynamic simulations, clumping on
the scale of individual star-forming regions is far below our
resolution and is not included. The most likely effect of clumping
would be to open up ``holes'' with lower attenuation, so that our
simulations would underestimate the frequency with which young stellar
populations are visible.

Another phenomenon which is treated poorly in our simulations is gas
outflows.  Starburst galaxies are almost ubiquitously observed to have
large-scale gas outflows 
\citep[e.g.][]{heckmanetal00}.
While our feedback prescription does provide energy input from
supernovae, it does not seem to lead to significant outflows. 
Furthermore, recent simulations including active galactic nuclei
\citep{springeletal04} show that AGN feedback can drive a powerful
outflow, clearing out gas from the central regions of the galaxies. 
This phenomenon is also not included in our present simulations. In the
context of dust attenuation, this lack of outflows has two
implications.  First, outflows, like dust patchiness, would tend to
open up holes in the dust distribution. Second, the lack of outflows
also means that the metals produced by the starburst remain in the
starburst region.  Since the optical depth is determined using a
constant dust-to-metal ratio, this will translate to a larger opacity
within the star-forming regions than would be expected if a significant
fraction of the metals were ejected.

Both of these limitations make it probable that our simulations
overestimate the amount of dust attenuation to some degree, especially
in the late stages after the galactic nuclei have merged.  Less clear,
however, is whether the correlations found here would be ruined.  For
example, while an AGN-driven outflow will clear out gas and dust from
the galaxy and hence decrease the dust attenuation, it also truncates
star formation.  Thus, while the evolution of the (stellar) luminosity
of the starburst would be altered by this phenomenon, the correlation
between dust attenuation and the luminosity, metallicity, etc., of the
starburst would not necessarily be altered.  Work is underway to
improve our model and include these effects.

\section{Discussion}

One of the remarkable aspects of
Figure~\ref{plot_luminosity_attenuation} is that the low-luminosity
stages of the simulations, which comprise both the initial, separate 
spiral galaxies and the final merger remnant, overlap. A priori, there
seems to be no reason to expect that the attenuation should remain
constant if two galaxies are merged into one. However, from the fit
results in Table~\ref{table-fitting-bolometric} it is evident that the
powers of the extensive quantities $L$, $\dot M_\star$, and $M_b$
approximately add to zero. This results in a dust attenuation which is
almost insensitive to simple size scaling of the galaxies; rather than
being determined by luminosity or star-formation rate alone, the dust
attenuation seems to be governed by ``specific luminosity'' and
``specific star-formation rate'', i.e., $L / M_b$ and $\dot M_\star /
M_b$.  This is fortunate for our analysis, since it means that not
treating the two galaxies in the initial stages of the merger
separately does not bias the results. This is only true in the specific
case being treated here, where the two galaxies are identical.  If the
two galaxies were not identical, as when simulating minor mergers or a
merger between a spiral and an elliptical galaxy, the dust attenuation
would have to be determined separately for the two components
regardless of whether $\alpha$, $\beta$, and $\gamma$ sum to zero.

It should, however, be pointed out that $\alpha + \beta + \gamma
\approx 0$ is a poor approximation for the fits to the attenuation in
the GALEX bands in Table~\ref{table-fitting-filters}. There is thus the
possibility that the fits in these bands have been biased by the use of
system, rather than individual galaxy, quantities.  Indeed, there is a
significant discrepancy between the fits and the initial stages of the
simulations in these bands, such that the fits overestimate the amount
of dust attenuation.

Looking at the parameter sets in Table~\ref{table-fitting-filters},
clear trends with wavelength can be seen. Going from ultraviolet to
near-infrared wavelengths, the dependence on star-formation rate
decreases while the dependence on luminosity increases.  This is not
unexpected; the luminosity in the ultraviolet is dominated by massive,
short-lived stars, and hence correlates well with star-formation rate. 
At longer wavelengths, contributions to the luminosity come from stars
of a wide range in age and is better represented by the bolometric
luminosity of the galaxy.  This trend with wavelength thus contains
information about the stars whose radiation is being absorbed.  What is
more surprising is that the trend is reversed at wavelengths longer
than $2 \um$.  Naively, one would expect that the dust attenuation at
progressively longer wavelengths always would be more dominated by
older stars, but this does not seem to be the case.  This effect
probably originates in the fact that around the age of $10 \Myr$, a
stellar population is very bright in the near-infrared due to the
presence of red supergiants. This means that young stellar populations
make a larger contribution to the emission at several $\um$ than what
is expected from their main-sequence temperatures.

It was earlier noted that, as shown in
Figure~\ref{plot_fiducial_luminosity}, the UV/visual luminosity not
absorbed by dust essentially is independent of the intrinsic bolometric
luminosity. This notion is confirmed by the fit to the bolometric
attenuation independent of the star-formation rate in
Table~\ref{table-fitting-bolometric}.  That fit yielded $\beta = 1.10$,
close to $\beta = 1$ for which the increase in luminosity  is exactly
compensated by the increase in attenuation given by
equation~\ref{equation-toymodel} (in the optically thick limit).

Is the fact that the dust attenuation in the simulations is well
described by equation~\ref{equation-fitting} indicating something about
the relative geometry of dust and stars in the simulations?  As already
mentioned, gas density is the driving factor behind both dust optical
depth and star-formation rate.  Furthermore, in the simulations, dust
and stars are assumed to be uniformly mixed within individual grid
cells.  It is thus not unreasonable to expect that a uniform mixture of
dust and stars should fit the simulations reasonably well.

The fact that the dust attenuation can be predicted so well by a simple
formula also implies that there are no hidden parameters determining
whether a luminous galaxy will be a ULIRG or not. 
\citet{bekkishioya00b}, from analyzing a prograde-prograde and a
retrograde-retrograde major merger at the time of highest
star-formation rate, drew the conclusion that retrograde mergers should
have stronger internal dust attenuation than prograde ones. This led
them to conjecture that interacting galaxies without long tidal tails
should be more prevalent among ULIRGs.  This result is not confirmed by
our much more extensive analysis.  While there is a tendency for
retrograde mergers to induce slightly more intense starbursts in our
simulations, and hence be more obscured according to our fitting
formula, they follow the same relation for dust attenuation as mergers
of any other geometry.  At least within the parameter space covered by
these simulations, there should be no such thing as a ``naked''
vigorously star-forming, $10^{ 12 } \Lsun$ system. While the
attenuation along some lines of sight might be smaller, the vast
majority of the bolometric luminosity should always be emerging as
infrared dust emission.

Comparing the studies of \citet{wangheckman96} and \citet{vijhetal03}
to these simulation results, it is encouraging to note that the
dependence of optical depth on luminosity is similar. Unfortunately,
they did not have metallicity or mass information, so they were unable
to perform a multi-dimensional fit like the one in this study.  It
would be interesting to know whether the dependence on these parameters
would also be similar to what is found here.

The fitting formulae presented here are in the ``theoretical plane'',
i.e., they depend on quantities such as total baryonic mass and
bolometric luminosity which are generally known in theoretical models
but hard to determine from observations.  As such, they are of limited
use for interpreting observations.  It has already been emphasized that
dust effects cancel the effects of increased luminosity to a remarkable
degree, so that the simulations show virtually no correlation between
the \emph{apparent} luminosity and the dust attenuation, either
bolometric or in the ultraviolet.  Unfortunately, this implies that
drawing conclusions about these highly dust-extinguished systems from
their apparent luminosity is very difficult, unless infrared data are
available. In particular, ``correcting'' apparent luminosities for dust
using a luminosity-based prescription is not likely to work well, a
fact also noted by \citet{hopkinsetal01}.

The relations presented here should be suitable for inclusion in
theoretical models for galaxy formation, such as semi-analytic models
(SAMs) including merger-driven starbursts
\citep[e.g.,][]{SPF}.
Unlike other theoretical models for dust effects
\citep[e.g.,][]{silvaetal98, charlotfall00}
our results give information about the \emph{magnitude} of the dust
attenuation and its dependence on the properties of the galaxy in a
self-consistent way.  Current approaches used to incorporate dust in
SAMs include relying on empirical results like those of WH96 or even
simpler approximations such as a uniform slab with optical depth
proportional to gas column density times metallicity (essentially the
toy model presented here).  Even in more sophisticated models, such as
those of \citet{granatoetal00}, the optical depth is typically
determined in the same simple way.  These models also contain numerous
adjustable parameters whose values are not given by the SAM and thus
must be fixed at constant values.  Given that our fits appear to work
well across a wide range of galaxy properties, including these results
in SAMs should lead to a more realistic estimation of the effects of
dust in cosmological scenarios.

Finally, nothing has been said here about the \emph{shape} of the dust
attenuation curve, e.g., if the simulated galaxies obey the ``Calzetti
law'' \citep{calzettietal94}.  This information is contained in our
fits, and a future paper will explore this in detail. For now, we
simply remark that our simulations are inconsistent with a simple
``screen-like'' attenuation curve like the Calzetti law, and do not
even follow any single reddening law resulting from more complicated
geometries \citep{wittgordon96, gcw97}. This is likely the result of a
stochastic superposition of many star-forming regions with various
optical depths and ages.

\section{Summary}

We have presented results from radiative-transfer calculations in a
comprehensive suite of galaxy merger simulations.  The results from
these simulations consist of images at many different wavelengths, as
well as spectral energy distributions, and seem to agree well with
observed properties of starburst galaxies.

It was discovered that the dust attenuation (defined as the fraction of
luminosity which is absorbed by dust) in the simulations can be
predicted from the bolometric luminosity, star-formation rate, baryonic
mass, and average gas metallicity of the system through a simple,
physically motivated formula (eq.~\ref{equation-fitting}).  Averaged
over all directions, the attenuation of the bolometric luminosity can
be predicted with a scatter of $4 \%$.  The attenuation along a
specific line of sight can be predicted with a scatter of 6 -- $12 \%$,
depending on wavelength.  The increased scatter is largely a result of
the variation of attenuation with viewing angle.  These relations are
valid for simulations with a range of two orders of magnitude in mass,
with metallicities from 0.3 to 1.1$\, \rm { Z }_\odot$, and gas
fractions from 20 to $60 \%$. The relations also seem to be valid for
both isolated and interacting galaxies.

Our results are consistent with studies of observed galaxies by
\citet{wangheckman96} and \citet{vijhetal03}, but our inclusion of
additional independent variables significantly lowers the scatter
around the relations, and we present results valid for wavelengths from
the far-ultraviolet to the near-infrared.

\acknowledgments
We thank Volker Springel for making GADGET and his initial-conditions
generator available to us.  We also thank Jennifer Lotz and Kathy
Cooksey for useful feedback on a draft version of this paper.  This
research used computational resources of the National Energy Research
Scientific Computing Center (NERSC), which is supported by the Office
of Science of the U.S. Department of Energy and also UpsAnd, a Beowulf
at UCSC. PJ was supported by a grant from IGPP/LLNL,  TJC and JRP by
grants from NASA and NSF.

\bibliographystyle{../../../bib/another-apj} 
\bibliography{../../../bib/patriks} \end {document}

%% file: natlatex_macros.tex
\newcommand{\nm}{\, {\rm nm}}

\newcommand{\um}{\, {\mu \rm m}}

\newcommand{\yr}{\, {\rm yr}}
\newcommand{\Myr}{\,\mathrm{Myr}}
\newcommand{\Gyr}{\,\mathrm{Gyr}}

\newcommand{\kpc}{\, {\rm kpc}}

\newcommand{\Angstrom}{\, {\rm \AA}}
\newcommand{\degrees}{^{\circ}}

\newcommand{\Msun}{\,\mathrm{M}_{\odot}}
\newcommand{\Lsun}{\,\mathrm{L}_{\odot}}

\providecommand{\halpha}{{\mathrm{H}\alpha}}
\providecommand{\hbeta}{{\mathrm{H}\beta}}
\renewcommand{\halpha}{{\mathrm{H}\alpha}}
\renewcommand{\hbeta}{{\mathrm{H}\beta}}
\newcommand{\mcrx}{\textit{Sunrise}}

%% file: attenuation_letter.bbl
\begin{thebibliography}{55}
\expandafter\ifx\csname natexlab\endcsname\relax\def\natexlab#1{#1}\fi

\bibitem[{{Adelberger} \& {Steidel}(2000)}]{adelbergersteidel00}
{Adelberger}, K.~L., \& {Steidel}, C.~C. 2000, \apj, 544, 218

\bibitem[{{Baldry} {et~al.}(2004){Baldry}, {Glazebrook}, {Brinkmann}, {Ivezi{\'
  c}}, {Lupton}, {Nichol}, \& {Szalay}}]{baldryetal04}
{Baldry}, I.~K., {Glazebrook}, K., {Brinkmann}, J., {Ivezi{\' c}}, {\v Z}.,
  {Lupton}, R.~H., {Nichol}, R.~C., \& {Szalay}, A.~S. 2004, \apj, 600, 681

\bibitem[{{Bekki} \& {Shioya}(2000{\natexlab{a}})}]{bekkishioya00a}
{Bekki}, K., \& {Shioya}, Y. 2000{\natexlab{a}}, \apj, 542, 201

\bibitem[{{Bekki} \& {Shioya}(2000{\natexlab{b}})}]{bekkishioya00b}
---. 2000{\natexlab{b}}, \aap, 362, 97

\bibitem[{{Bekki} \& {Shioya}(2001)}]{bekkishioya01}
---. 2001, \apjs, 134, 241

\bibitem[{{Bell} \& {Kennicutt}(2001)}]{bellkennicutt01}
{Bell}, E.~F., \& {Kennicutt}, R.~C. 2001, \apj, 548, 681

\bibitem[{{Bell} {et~al.}(2003){Bell}, {McIntosh}, {Katz}, \&
  {Weinberg}}]{belletal03}
{Bell}, E.~F., {McIntosh}, D.~H., {Katz}, N., \& {Weinberg}, M.~D. 2003, \apjl,
  585, L117

\bibitem[{{Buat} {et~al.}(2004){Buat}, {Iglesias-Paramo}, {Seibert},
  {Burgarella}, {Charlot}, {Martin}, {Xu}, {Heckman}, \&
  {Boissier}}]{buatetal04}
{Buat}, V. {et~al.} 2004, \apjs, in press (astro-ph/0411343)

\bibitem[{{Calzetti} {et~al.}(1994){Calzetti}, {Kinney}, \&
  {Storchi-Bergmann}}]{calzettietal94}
{Calzetti}, D., {Kinney}, A.~L., \& {Storchi-Bergmann}, T. 1994, \apj, 429, 582

\bibitem[{{Charlot} \& {Fall}(2000)}]{charlotfall00}
{Charlot}, S., \& {Fall}, S.~M. 2000, \apj, 539, 718

\bibitem[{Cox(2004)}]{tjthesis-nourl}
Cox, T.~J. 2004, PhD thesis, UC Santa Cruz

\bibitem[{{Cox} {et~al.}(2005){Cox}, {Jonsson}, {Primack}, \&
  {Somerville}}]{coxetal05methods}
{Cox}, T.~J., {Jonsson}, P., {Primack}, J., \& {Somerville}, R.~S. 2005,
  \mnras, submitted

\bibitem[{{Cox} {et~al.}(2004){Cox}, {Primack}, {Jonsson}, \&
  {Somerville}}]{coxetal04}
{Cox}, T.~J., {Primack}, J., {Jonsson}, P., \& {Somerville}, R.~S. 2004, \apjl,
  607, L87

\bibitem[{{de Jong}(1996)}]{dejong96}
{de Jong}, R.~S. 1996, \aap, 313, 45

\bibitem[{{Devriendt} {et~al.}(1999){Devriendt}, {Guiderdoni}, \&
  {Sadat}}]{devriendtetal99}
{Devriendt}, J.~E.~G., {Guiderdoni}, B., \& {Sadat}, R. 1999, \aap, 350, 381

\bibitem[{{Dwek}(1998)}]{dwek98}
{Dwek}, E. 1998, \apj, 501, 643

\bibitem[{{Elbaz} \& {Cesarsky}(2003)}]{elbazcesarsky03}
{Elbaz}, D., \& {Cesarsky}, C.~J. 2003, Science, 300, 270

\bibitem[{{Ferrara} {et~al.}(1999){Ferrara}, {Bianchi}, {Cimatti}, \&
  {Giovanardi}}]{ferraraetal99}
{Ferrara}, A., {Bianchi}, S., {Cimatti}, A., \& {Giovanardi}, C. 1999, \apjs,
  123, 437

\bibitem[{{Fukugita} {et~al.}(1996){Fukugita}, {Ichikawa}, {Gunn}, {Doi},
  {Shimasaku}, \& {Schneider}}]{fukugitaetal96}
{Fukugita}, M., {Ichikawa}, T., {Gunn}, J.~E., {Doi}, M., {Shimasaku}, K., \&
  {Schneider}, D.~P. 1996, \aj, 111, 1748

\bibitem[{{Goldader} {et~al.}(2002){Goldader}, {Meurer}, {Heckman}, {Seibert},
  {Sanders}, {Calzetti}, \& {Steidel}}]{goldaderetal02}
{Goldader}, J.~D., {Meurer}, G., {Heckman}, T.~M., {Seibert}, M., {Sanders},
  D.~B., {Calzetti}, D., \& {Steidel}, C.~C. 2002, \apj, 568, 651

\bibitem[{{Gordon} {et~al.}(1997){Gordon}, {Calzetti}, \& {Witt}}]{gcw97}
{Gordon}, K.~D., {Calzetti}, D., \& {Witt}, A.~N. 1997, \apj, 487, 625

\bibitem[{{Gordon} {et~al.}(2001){Gordon}, {Misselt}, {Witt}, \&
  {Clayton}}]{gordonetal01}
{Gordon}, K.~D., {Misselt}, K.~A., {Witt}, A.~N., \& {Clayton}, G.~C. 2001,
  \apj, 551, 269

\bibitem[{{Granato} {et~al.}(2000){Granato}, {Lacey}, {Silva}, {Bressan},
  {Baugh}, {Cole}, \& {Frenk}}]{granatoetal00}
{Granato}, G.~L., {Lacey}, C.~G., {Silva}, L., {Bressan}, A., {Baugh}, C.~M.,
  {Cole}, S., \& {Frenk}, C.~S. 2000, \apj, 542, 710

\bibitem[{{Heckman} {et~al.}(2000){Heckman}, {Lehnert}, {Strickland}, \&
  {Armus}}]{heckmanetal00}
{Heckman}, T.~M., {Lehnert}, M.~D., {Strickland}, D.~K., \& {Armus}, L. 2000,
  \apjs, 129, 493

\bibitem[{{Heckman} {et~al.}(1998){Heckman}, {Robert}, {Leitherer}, {Garnett},
  \& {van der Rydt}}]{h98}
{Heckman}, T.~M., {Robert}, C., {Leitherer}, C., {Garnett}, D.~R., \& {van der
  Rydt}, F. 1998, \apj, 503, 646

\bibitem[{{Hopkins} {et~al.}(2001){Hopkins}, {Connolly}, {Haarsma}, \&
  {Cram}}]{hopkinsetal01}
{Hopkins}, A.~M., {Connolly}, A.~J., {Haarsma}, D.~B., \& {Cram}, L.~E. 2001,
  \aj, 122, 288

\bibitem[{{Jonsson}(2004)}]{pjthesis-nourl}
{Jonsson}, P. 2004, PhD thesis, UC Santa Cruz

\bibitem[{{Kroupa}(2002)}]{kroupa02}
{Kroupa}, P. 2002, in ASP Conf. Ser., ed. E.~K. {Grebel} \& W.~{Brandner}, Vol.
  285 (San Francisco: ASP), 86--

\bibitem[{{Leitherer} {et~al.}(1999){Leitherer}, {Schaerer}, {Goldader},
  {Delgado}, {Robert}, {Kune}, {de Mello}, {Devost}, \&
  {Heckman}}]{leithereretal99}
{Leitherer}, C. {et~al.} 1999, \apjs, 123, 3

\bibitem[{{Lotz} {et~al.}(2004){Lotz}, {Primack}, \& {Madau}}]{lotzetal04gm20}
{Lotz}, J.~M., {Primack}, J., \& {Madau}, P. 2004, \aj, 128, 163

\bibitem[{{Lupton} {et~al.}(2003){Lupton}, {Blanton}, {Fekete}, {Hogg},
  {O'Mullane}, {Szalay}, \& {Wherry}}]{lupton03}
{Lupton}, R., {Blanton}, M.~R., {Fekete}, G., {Hogg}, D.~W., {O'Mullane}, W.,
  {Szalay}, A., \& {Wherry}, N. 2003, astro-ph/0312483

\bibitem[{{Meurer} {et~al.}(1999){Meurer}, {Heckman}, \& {Calzetti}}]{mhc99}
{Meurer}, G.~R., {Heckman}, T.~M., \& {Calzetti}, D. 1999, \apj, 521, 64

\bibitem[{{Mihos} \& {Hernquist}(1994)}]{mihoshernquist94ulirgs}
{Mihos}, J.~C., \& {Hernquist}, L. 1994, \apjl, 431, L9

\bibitem[{{Mihos} \& {Hernquist}(1996)}]{mihoshernquist96}
---. 1996, \apj, 464, 641

\bibitem[{{Roberts} \& {Haynes}(1994)}]{robertshaynes94}
{Roberts}, M.~S., \& {Haynes}, M.~P. 1994, \araa, 32, 115

\bibitem[{{Sanders} \& {Mirabel}(1996)}]{sandersmirabel96}
{Sanders}, D.~B., \& {Mirabel}, I.~F. 1996, \araa, 34, 749

\bibitem[{{Schwarz}(1981)}]{schwarz81}
{Schwarz}, M.~P. 1981, \apj, 247, 77

\bibitem[{{Shen} {et~al.}(2003){Shen}, {Mo}, {White}, {Blanton}, {Kauffmann},
  {Voges}, {Brinkmann}, \& {Csabai}}]{shenetal03}
{Shen}, S., {Mo}, H.~J., {White}, S.~D.~M., {Blanton}, M.~R., {Kauffmann}, G.,
  {Voges}, W., {Brinkmann}, J., \& {Csabai}, I. 2003, \mnras, 343, 978

\bibitem[{{Silva} {et~al.}(1998){Silva}, {Granato}, {Bressan}, \&
  {Danese}}]{silvaetal98}
{Silva}, L., {Granato}, G.~L., {Bressan}, A., \& {Danese}, L. 1998, \apj, 509,
  103

\bibitem[{{Smail} {et~al.}(1997){Smail}, {Ivison}, \& {Blain}}]{smailetal97}
{Smail}, I., {Ivison}, R.~J., \& {Blain}, A.~W. 1997, \apjl, 490, L5

\bibitem[{{Somerville} {et~al.}(2001){Somerville}, {Primack}, \& {Faber}}]{SPF}
{Somerville}, R.~S., {Primack}, J.~R., \& {Faber}, S.~M. 2001, \mnras, 320, 504

\bibitem[{{Springel}(2000)}]{springel00}
{Springel}, V. 2000, \mnras, 312, 859

\bibitem[{{Springel} {et~al.}(2005{\natexlab{a}}){Springel}, {Di Matteo}, \&
  {Hernquist}}]{springeletal04agnb}
{Springel}, V., {Di Matteo}, T., \& {Hernquist}, L. 2005{\natexlab{a}}, ApJL,
  in press (astro-ph/0409436)

\bibitem[{{Springel} {et~al.}(2005{\natexlab{b}}){Springel}, {Di Matteo}, \&
  {Hernquist}}]{springeletal04}
---. 2005{\natexlab{b}}, \mnras, submitted (astro-ph/0411108)

\bibitem[{{Springel} \& {Hernquist}(2002)}]{springelhernquist02}
{Springel}, V., \& {Hernquist}, L. 2002, \mnras, 333, 649

\bibitem[{{Springel} \& {Hernquist}(2005)}]{springelhernquist04}
---. 2005, ApJL, submitted (astro-ph/0411379)

\bibitem[{{Springel} {et~al.}(2001){Springel}, {Yoshida}, \&
  {White}}]{springeletal01}
{Springel}, V., {Yoshida}, N., \& {White}, S.~D.~M. 2001, New Astronomy, 6, 79

\bibitem[{{Sullivan} {et~al.}(2001){Sullivan}, {Mobasher}, {Chan}, {Cram},
  {Ellis}, {Treyer}, \& {Hopkins}}]{sullivanetal01}
{Sullivan}, M., {Mobasher}, B., {Chan}, B., {Cram}, L., {Ellis}, R., {Treyer},
  M., \& {Hopkins}, A. 2001, \apj, 558, 72

\bibitem[{{Vijh} {et~al.}(2003){Vijh}, {Witt}, \& {Gordon}}]{vijhetal03}
{Vijh}, U.~P., {Witt}, A.~N., \& {Gordon}, K.~D. 2003, \apj, 587, 533

\bibitem[{{Wang} \& {Heckman}(1996)}]{wangheckman96}
{Wang}, B., \& {Heckman}, T.~M. 1996, \apj, 457, 645

\bibitem[{{Weingartner} \& {Draine}(2001)}]{weingartnerdraine01}
{Weingartner}, J.~C., \& {Draine}, B.~T. 2001, \apj, 548, 296

\bibitem[{{Witt} \& {Gordon}(1996)}]{wittgordon96}
{Witt}, A.~N., \& {Gordon}, K.~D. 1996, \apj, 463, 681

\bibitem[{{Witt} \& {Gordon}(2000)}]{wittgordon00}
---. 2000, \apj, 528, 799

\bibitem[{{Wolf} {et~al.}(2004){Wolf}, {Bell}, {McIntosh}, {Rix}, {Barden},
  {Beckwith}, {Borch}, {Caldwell}, {Haeussler}, {Heymans}, {Jahnke}, {Jogee},
  {Meisenheimer}, {Peng}, {Sanchez}, {Somerville}, \& {Wisotzki}}]{wolf04}
{Wolf}, C. {et~al.} 2004, \apj, submitted (astro-ph/0408289)

\bibitem[{{Zaritsky} {et~al.}(1994){Zaritsky}, {Kennicutt}, \&
  {Huchra}}]{zaritskyetal94}
{Zaritsky}, D., {Kennicutt}, R.~C., \& {Huchra}, J.~P. 1994, \apj, 420, 87

\end{thebibliography}
